\documentclass[journal]{IEEEtran}
\IEEEoverridecommandlockouts
\makeatletter
\def\ps@headings{%
\def\@oddhead{\mbox{}\scriptsize\rightmark \hfil \thepage}%
\def\@evenhead{\scriptsize\thepage \hfil \leftmark\mbox{}}%
\def\@oddfoot{}%
\def\@evenfoot{}}
\makeatother
\usepackage{subfigure}
\usepackage{colortbl}
\usepackage{bm}

\usepackage{graphicx}  
\usepackage{url}       

\usepackage{amsmath}   
\usepackage{cite}

\usepackage{amsfonts,amssymb}

\usepackage{stfloats}

\usepackage{cases}
\usepackage{algorithm}
\usepackage{multirow}
\usepackage{algorithmic}
\usepackage{stfloats}
\usepackage{epstopdf}
\usepackage{esint}

\usepackage{xcolor}
\usepackage{xpatch}
\makeatletter
\def\changeBibColor#1{%
	\in@{#1}{THz_RIS, discreteRIS3, THZ_estimation, Gaussian_Error,Pilot_pollution,Star_RIS,ETSI,3GPP_16}
	\ifin@\else\normalcolor\fi
}
\xpatchcmd\@bibitem
{\item}
{\changeBibColor{#1}\item}
{}{\fail}
\xpatchcmd\@lbibitem
{\item}
{\changeBibColor{#2}\item}
{}{\fail}

\newcommand{\Rmnum}[1]{\expandafter\@slowromancap\romannumeral #1@}

\makeatother

\newtheorem{theorem}{{Theorem}}

\newtheorem{remark}{{Remark}}

\newcommand{\ls}[1]
    {\dimen0=\fontdimen6\the\font
     \lineskip=#1\dimen0
     \advance\lineskip.5\fontdimen5\the\font
     \advance\lineskip-\dimen0
     \lineskiplimit=.9\lineskip
     \baselineskip=\lineskip
     \advance\baselineskip\dimen0
     \normallineskip\lineskip
     \normallineskiplimit\lineskiplimit
     \normalbaselineskip\baselineskip
     \ignorespaces
    }

\begin{document}

\title{Huygens-Fresnel Model Based \\Position-Aided Phase Configuration for \\1-Bit RIS Assisted Wireless Communication}
\vspace{10pt}

\author{
\IEEEauthorblockN{Xiao Zheng, \emph{Graduate Student Member}, \emph{IEEE}, Wenchi Cheng, \emph{Senior Member}, \emph{IEEE}, and Jiangzhou Wang, \emph{Fellow}, \emph{IEEE}}


\thanks{Part of this work was presented in IEEE Global Communications conference, 2022\cite{xiaozheng}.

This work was supported in part by the National
Natural Science Foundation of China (No.
62341132), the National Key Research and
Development Program of China under Grant
2021YFC3002102, and the Key R\&D Plan of
Shaanxi Province under Grant 2022ZDLGY05-09.

Xiao Zheng and Wenchi Cheng are with the State Key Laboratory of Integrated Services Networks, Xidian University, Xi'an, 710071, China (e-mails: zheng\_xiao@stu.xidian.edu.cn, wccheng@xidian.edu.cn).

Jiangzhou Wang is with the School of Engineering, the University of Kent, Canterbury, United Kingdom (e-mail: j.z.wang@kent.ac.uk).

}
}

\maketitle

\begin{abstract}
Reconfigurable intelligent surface (RIS), composed of nearly passive elements, is regarded as one of the potential paradigms to support multi-gigabit data in real-time. However, in traditional CSI (channel state information) driven frame, the training overhead of channel estimation greatly increases as the number of RIS elements increases to intelligently manipulate the reflected signals. To conveniently use the reflected signal without complex CSI feedback, in this paper we propose a position-aided phase configuration scheme based on the property of Fresnel zone. In particular, we design the impedance based discrete RIS elements with joint absorption mode and reflection mode considering the fabrication complexities, which integrated the property of the Fresnel zone to resist the impact of position error. Then, with joint absorption and 1-bit reflection mode elements, we develop the two-step position-aided ON/OFF states judgement (TPOSJ) scheme and the frame structure to control the ON/OFF state of RIS, followed by analyzing the impacts of mobility and position error on our proposed scheme. Also, we derive the Helmholtz-Kirchhoff integral theorem based power flow. Simulations show that the proposed scheme can manipulate the ON/OFF state intelligently without complex CSI, thus verifying the practical application of our proposed scheme.
\end{abstract}


\section{Introduction}
\IEEEPARstart{O}{ver} the past few years, the urgent demands for ultra-high speed and low-latency have experienced rapid growth with the explosive increase of mobile devices and burgeoning applications\cite{5G,Yixin,Wenchi}. In view of these demands, operating at much higher carrier frequencies (within the range from 30 GHz to 10 THz) in wireless environments is essential since there are thousands of times of bandwidth compared with 4G frequency band\cite{high,wang1,wang2}. However, operating in high frequency bands undergoes severe propagation attenuation, which reduces the coverage area and suffers from obstructions\cite{mmWave}. It is noteworthy that reconfigurable intelligent surface (RIS), which is composed of passive elements and leveraged to provide additional reflecting links, can be conveniently applied to remedy the deficiency of high propagation loss\cite{Haiyue,Magazine1,bao}. Designed by metamaterial, the passive and reconfigurable elements can manipulate both the amplitude and phase of the incident signal to meet the demands of various services. More importantly, compared to an active relay, the power consumption of RIS is negligible since it does not require radio frequency chains, which is energy-intensive. Combining RIS with high-frequency bands is regarded as one of the promising approaches to enable ultra-high data rate and low-latency communications\cite{Magazine2, china1}. 

However, as the intensity of reflected signal is proportional to the square of the RIS area according to power scaling law\cite{RIS-size}, the number of RIS elements increases as radio frequency increases when the size of RIS is fixed, considering the design criteria that adjacent elements spacing of sub-wavelength scale to avoid coupling effect. Consequently, new problems occur in two aspects: Firstly, the plane wave hypothesis is no longer supported since the transmitter and receiver are generally situated in the near-field region of RIS array as radio frequency increases. Under this circumstance, existing schemes based on far-field propagation model are no longer valid. Secondly, the channel estimation needs to be considered in RIS-assisted communications since the channel state information (CSI) of RIS-associated link should be known in advance to design optimum phase shift. However, the training overhead of RIS-associated links is enormous, especially for large size RIS.

To be specific, the near-field distance (i.e. Fraunhofer distance) is a critical point to distinguish spherical wave and planar wave and will increase as the radio frequency and the size of RIS array rise. The Fraunhofer distance in millimeter wave (mmWave) bands is of hundred meters scale when the RIS area in square shape is 1 $\rm m^2$, which nearly covers the radius of a typical 5G cell, not to mention in higher frequency bands\cite{Magazine3}. {In near-field region, the received signal from RIS exposes a seemingly oscillatory behavior for the planar and linear configurations, respectively\cite{RISFresnel}. Thus, the addition of range information is of great importance in the near-field region.  Without proper design, the intensity of received signal is likely to be at a trough, and the expected gain induced by RIS cannot be achieved. Under this circumstance, exploiting spherical wave propagation model is necessary while some schemes based on classical far-field approximation are not applicable. To comprehend the operation mechanism of RIS profoundly, some papers investigated the near-field properties of RIS, which is summarized in TABLE I.} The latter issue about estimation overhead of RIS-assisted communications is one of the inherent problems for RIS with large aperture\cite{THZ_estimation}. {In traditional CSI-driven RIS-assisted communications, the CSI estimation is based on orthogonal pilot symbols. The training overhead of full CSI scales at least linearly with the number of RIS elements or transmit/receive antenna, which will make the intelligent phase manipulation unattainable, especially for the large RIS in high frequency. Meanwhile, the deployment of RIS will greatly increase the complexity of channel estimation as the link from Tx to RIS and the link from RIS to Rx should be estimated separately \cite{RISestimation1,china2}. Under this condition, the performance of RIS-assisted communications is seriously limited.} Some authors attempted to reduce the cost of frequent channel estimation from various aspects. However, when operating in the near-field, the channel shows spatial non-stationarity and is even more informative\cite{Holographic-RIS}, making the channel estimation problem more complicated. 

\begin{table*}[t!] 
    \begin{center}
    \vspace{-0pt}
    {\caption{Related works about  electromagnetic properties of RIS in the near-field}}
    \label{I}
    \begin{tabular}{| m{0.8cm}<{\centering}|  m{8.5cm}<{\centering}|  m{2.5cm}<{\centering}|  m{2.5cm}<{\centering}|}
      \hline
      {\textbf{Ref.}} & {\textbf{Main Idea}} &  {\textbf{Foundation} }& {\textbf{Operating Region}} \\
      \hline
      {\cite{RISFresnel}} & {The Fresnel zone decomposition was proposed to unify the scattering and reflection of RIS. }& {Electromagnetic theory }& {Near-field and far-field }\\\hline
      {\cite{Near-Field1} }&{Two metrics of
      benefit distance and near-field gain were defined to characterize the
      electric fields in the near-field region.}  &{Ray optics}    & {Near-field} \\\hline
      {\cite{NearfiledRIS3} } &{A scheme of RIS's coefficient design was developed to leverage the power gain of spherical and cylindrical waves.} &{Electromagnetic theory } &{Near-field} \\\hline
      {\cite{RIS_Path_Loss}}  &{The free-space path loss models in both near-field and far-field was proposed, which is validated by measurement.} &{Electromagnetic theory}  &{Near-field and far-field } \\\hline
      {\cite{Green_Theorem}} &{A physics-consistent analytical characterization of the free-space path-loss for a RIS-assisted wireless link was proposed.}  &{Green's theorem } &{Near-field}  \\\hline
      {\cite{Electromagnetic_Model}} &{An accurate and simple analytical model was proposed for the computation of the reflection amplitude and phase of RIS.}  &{ Radar cross section} &{Near-field} \\\hline
\end{tabular}
\end{center}
\vspace{-15pt}
\end{table*}

{Nevertheless, due to the energy-efficient and cost-effective nature of RIS, it is a trend to deploy extremely large size RIS in future 6G communications (called holographic communications \cite{Holographic-RIS}). Integrating the actuality that the operational frequency in wireless communications moves on higher bands, the transmission environments will be dominated by the virtual LoS propagation paths reflected by RIS. The LoS probability greatly increases as the size of RIS increases, exposing the same behavior as the near-field distance \cite{RISestimation1}. That is, the channel in future RIS-assisted communications shows a highly geometry-related property. Thus, the position information has been pointed out to provides valuable prior side-information for RIS-assisted communications.} The authors in \cite{location-aid1} investigated the multi-RIS-assisted system from a dual perspective, which showed a tight interplay between localization and communications. The authors in \cite{location-aid4} verified that the introduction of prior location information can further accelerate the beam alignment and channel parameter estimation processes. The authors in \cite{location-aid2} proposed a joint RIS location and passive beamforming optimization scheme for physical layer security. The authors in \cite{location-aid3} proposed a location information assisted beamforming design without the requirement of the channel training process. {Compared with traditional CSI-driven communications, the position-aided schemes have the potentials to decrease the overall complexity and capture the mobility in RIS-assisted communications, providing valuable side-information about the physical layer\cite{location_aware}. However, designing a complete position-driven scheme for RIS-assisted communications has not been addressed in the aforementioned works.}

Motivated by the above observations, to tackle the problem of high estimation overhead in traditional CSI-driven RIS-assisted communications, in this paper we propose the CSI-free phase configuration scheme considering the position error in practical scenarios and design the corresponding frame, followed by the power flow. The main contributions of this paper are summarized as follows:
{
\begin{itemize}
\item Considering the hard attainment of real-time CSI in RIS-assisted near-field communications, we unite the position information with the Fresnel zone model in the near-field and propose the two-step position-aid ON/OFF states judgement (TPOSJ) scheme without rigid demand of CSI. Furthermore, we design the judgement threshold to achieve the tradeoff between achievable rate and robustness against position error. We show that exploiting the property of the Fresnel zone can achieve better robustness against position error compared with \cite{RIS_Path_Loss}. 
\item To further improve the robustness, we design discrete RIS elements with joint absorption and reflection modes based on the ON/OFF functions of positive-intrinsic-negative (PIN) diodes, where the new element in absorption mode can completely absorb the incident wave without reflection. To our best knowledge, using complete absorption mode of element to improve communication performance is addressed for the first time in this paper. The absorption mode is proposed to resist severe performance degradation caused by position error, which is verified by simulations.
\item We evaluate the feasibility of proposed scheme in complexity analysis including computational complexity, estimation complexity and singling overhead, and analyse the impacts of mobility and position error. In particular, complexity analysis shows that the computational complexity of the TPOSJ scheme just increases linearly as the number of RIS elements increases. Furthermore, it has lower estimation complexity and singling overhead compared with CSI-driven scheme. Additionally, it offers the potential to support the Rx with high mobility and the impact of position error on the TPOSJ scheme is similar to additive white gaussian noise (AWGN) channel. Numerical results demonstrate that with joint absorption mode and reflection mode of RIS element, our proposed scheme is robust to position error to some extent, which is consistent with our derived results.
\item With the proposed TPOSJ scheme, to enable CSI-free power and rate controls, we develop the reflected power of RIS based on Huygens-Fresnel model and Helmholtz-Kirchhoff integral theorem in both LoS and NLoS cases. We show that the existing typical works derived by electromagnetic theory \cite{RIS_Path_Loss} and RCS \cite{Electromagnetic_Model} are special cases of our derived expressions. Also, exploiting such expressions, the performance of the TPOSJ scheme can be further facilitated by rotating the facing direction of RIS.
\end{itemize}
}

\section{System Model}
\begin{figure*}[t]
    \centering 
    \centerline{\includegraphics[width=14cm]{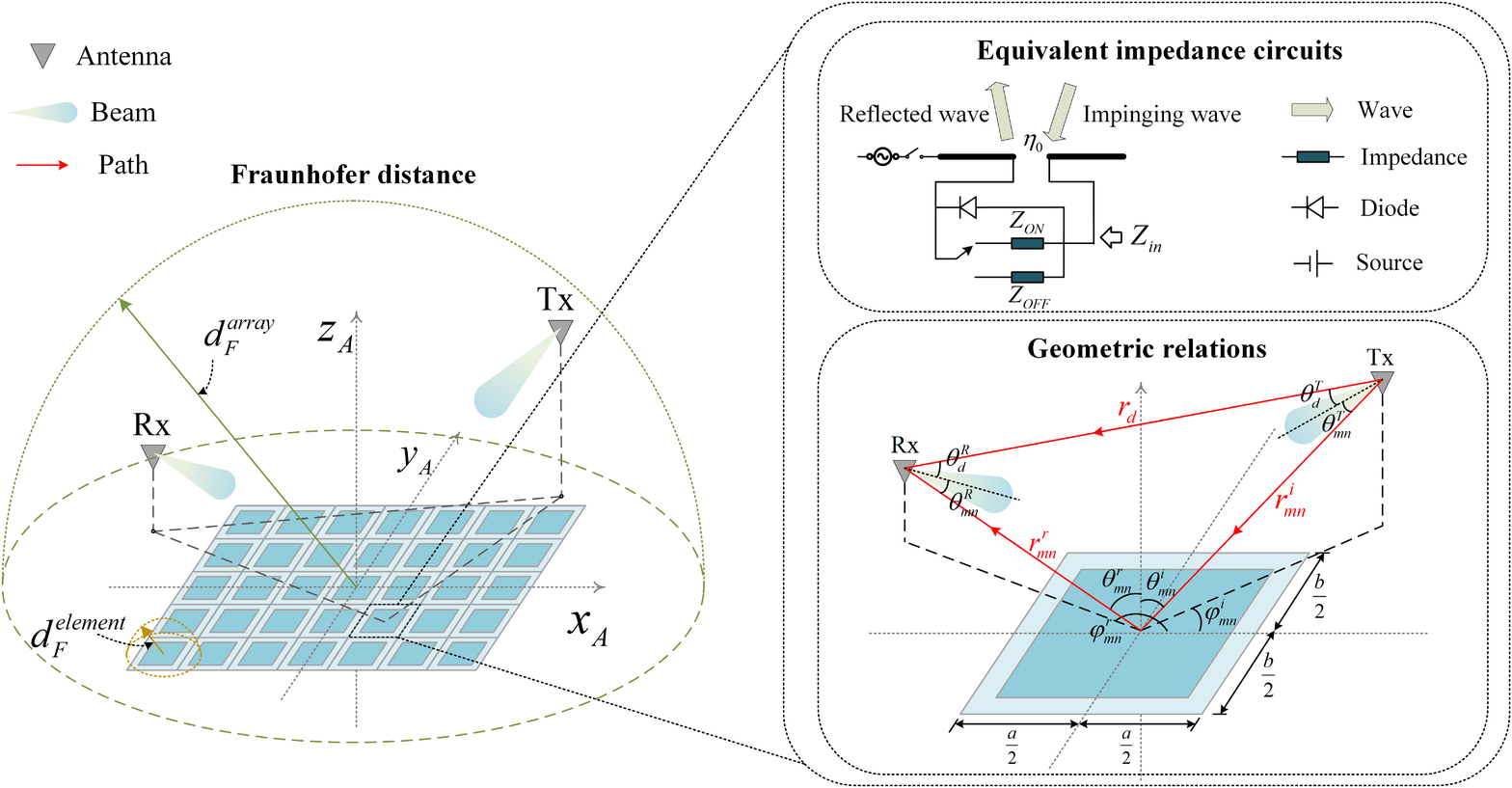}}
    \vspace{0pt}
    \caption{{The system model of RIS-assisted wireless communication.}}
    \label{system_model}
    \vspace{-10pt}
\end{figure*}
{The remainder of this paper is organized as follows. Section II introduces the system model including the Fraunhofer distance, impedance based elements, and geometric relations. In Section III, we combine the property of Fresnel zone with geometric relations to develop the CSI-free phase configuration scheme. Based on this, we design the frame structure. In Section IV, with our proposed scheme, we derive the power flow of RIS array in near-field region based on Helmholtz-Kirchhoff integral theorem. Section V evaluates the performance and feasibility of our proposed scheme as well as the impact of position error on spectrum efficiency. Conclusions are drawn in Section VI.}


We consider a general RIS-assisted wireless communication system as shown in Fig.~\ref{system_model}, where Tx and Rx represent the transmitter and receiver, respectively. The transmission is performed within the near-field region of RIS. In this section, we introduce the system model including near-field region, impedance based element, and geometric relations. 

\subsection{Fraunhofer Distance of Element and Array}
To leverage the property in the near-field regime, we need to know the principle of RIS's operating regime first. {The Fraunhofer distance, which is the critical point between near-field and far-field, is defined by considering the phase difference between the center and the corner caused by the wave's curvature. The maximum allowable phase difference in the far-field is $\pi$/8 while the largest phase variation occurs when the wave impinges perpendicularly. Thus, as shown in the left part of Fig.~\ref{system_model}, the Fraunhofer distances of RIS element and array, denoted by $d_F^{ element}$ and $d_F^{ array}$, respectively, can be expressed by 
\begin{align}
        d_F^{ element}=\frac{2\pi {D_{max}^{e}} ^2}{\lambda}=\frac{2\pi {D_{max}^{e}}^2f}{c}
\label{Fraunhofer1}
\end{align}
and
\begin{align}
    d_F^{ array}=\frac{2\pi {D_{max}^{a}}^2}{\lambda}=\frac{2\pi {D_{max}^{a}}^2f}{c},
\label{Fraunhofer2}
\end{align}
respectively, where $D_{max}^{e}$ and $D_{max}^{a}$ are the maximum lengths of element and array, respectively, $c$ is the speed of light, and $f$ is the radio frequency\cite{RISFresnel}. }

Based on Eqs.~\eqref{Fraunhofer1} and \eqref{Fraunhofer2}, the Fraunhofer distance is proportional to the radio frequency when the size of the target is fixed. Since the physical size of individual RIS element is usually of sub-wavelength scale, e.g., with the range from $\lambda$/10 to $\lambda$/2\cite{RIS_Path_Loss}, $d_F^{ element}$ is within the range from $\lambda$/50 to $\lambda$/2 when the RIS element is in a square. {The general communication distance is far greater than the Fraunhofer distance of individual element.} Thus, for individual element, the planar wave approximation is still valid. However, in mmWave bands, $d_F^{ array}$ is within hundreds of meters\cite{Holographic-RIS}, generally covering the radius of a typical 5G cell. The Tx and Rx will be situated in the near-field region of RIS array. Thus, in this paper the Tx/Rx are considered to be in the near-field of the RIS array and the far-field of individual RIS element. 

\subsection{Impedance Based 1-bit Element}
Considering the fabrication complexities and the material performance drop, only a limited number of discrete phase shifts can be achieved by a finite-sized RIS\cite{discreteRIS3}. One efficient approach to realize the function of discrete phase shift is by leveraging the switchable metamaterial, e.g., PIN diodes. Under impedance based model, the discrete phase shift can be achieved by different equivalent circuits resulting from the ON/OFF states of PIN diodes\cite{Electromagnetic_Model}. The reflected coefficient of the $k$th RIS element, denoted by ${\mit \Gamma}_k$, can be expressed as follows:
\begin{align}
    {\mit \Gamma} _k=\frac{Z_{ in}-\eta_0}{Z_{ in}+\eta_0}= {\mit \Gamma}  e^{j\psi_k}, \label{coefi}
\end{align}
where $Z_{ in}$ is the equivalent input impedance of RIS element, $\psi_k$ is the phase shift induced by the $k$th element, and ${\mit \Gamma}$ is the amplitude of the reflected coefficient. {${\mit \Gamma}$ satisfies ${\mit \Gamma}\in \left\{{\mit \Gamma}_{r},{\mit \Gamma}_{a}\right\}$ where ${\mit \Gamma}_{r}$ and ${\mit \Gamma}_{a}$ denote the reflected amplitudes of reflection mode and absorption mode, respectively.}{ The new RIS element with joint reflection and absorption modes can be achieved by the ON/OFF functions of PIN diodes as follows:
\begin{align}
    Z_{ in}=
    \begin{cases}
        Z_{ ON},~~~~\text{ reflection mode: ON};\\
        Z_{ OFF}, ~~~\text{reflection mode: OFF};\\
        \eta_0,~~~~~~~~\text{absorption mode},
    \end{cases} \label{Z_in}
\end{align}
where $\eta_0$ is the characteristic impedance of free space, $ Z_{ ON}$ and $Z_{ OFF}$ are the equivalent input impedances in reflection mode with ON state and OFF state, respectively. The reflection elements can manipulate the incident wave with a additional phase delay, where in this paper the phase shifts of ON and OFF states are designed as $\pi$ and 0, respectively. For the absorption elements, their coefficients satisfy ${\mit \Gamma}_{a}=0$ when the input impedance is $\eta_0$ based on Eqs.~\eqref{coefi} and \eqref{Z_in}, i.e., the element in absorption mode achieves impedance matching. In this case, the incident power is completely consumed by absorption elements\footnote{It is noteworthy that the elements with absorption mode are also passive. Its purpose is to attenuate the wireless incident wave, as opposed to requiring additional circuit power for amplifying the incident wave, as seen in active RIS.}. }


As shown in the top right corner of Fig.~\ref{system_model}, the proposed circuit of the element with joint absorption and reflection modes is given. The control of element needs two steps. {First, based on the state of the switch near the power source, the RIS element is in reflection mode when the power supply is on whereas the RIS element is in absorption mode with $Z_{ in}=\eta_0$ when the power supply is off, which needs one bit of control information. Then, when the RIS element is in reflection mode, the different equivalent circuits with $Z_{ in}=Z_{ ON}/Z_{ OFF}$, respectively, can be controlled by another bit, corresponding to different phase shifts. Hence, according to 2-bit of control information, the element can be adjusted as reflection elements with 1-bit phase delay resolution, or absorption elements to vanish the incident wave. The new elements are used to alleviate the impact of position error on the performance of the proposed scheme, which will be introduced in Section III.}

\subsection{Geometric Relations}
Figure~\ref{system_model} displays the geometry of a general RIS-assisted wireless communication system in the 3-dimensional Cartesian coordinate system. {We begin by considering the single-input single-output (SISO) case (arbitrary one of the pairs comprise the Rx antenna and one of the Tx antennas), and show that the scheme can be easily extended to multiple-input single-output (MISO) and multiple RISs case in the Section III. We randomly select one of the antennas of the Tx and denote by ${\rm T}^A(x_t^A, y_t^A, z_t^A)$ the coordinate of it in the $A$ coordinate system. We denote by ${\rm R}^A(x_r^A, y_r^A, z_r^A)$ the coordinate of Rx in the $A$ coordinate system.} The RIS is placed in the x-y plane and the center of RIS is located at the origin of the $A$ coordinate system. $N$ and $M$ denote the numbers of rows and columns of regularly arranged unit elements, respectively. The elements are assumed to be a rectangle, and the size of each element along the $x$-axis is $a$ and that along the $y$-axis is $b$. We denote by ${\rm P}_{mn}^A(P_{mn,x}^A, P_{mn,y}^A, P_{mn,z}^A)$ the coordinate point of the RIS element at the $m$th row and $n$th column, where $P_{mn,x}^A=\left(\frac{2m-M-1}{2}\right)a$, $m= 1,2,\cdots,M$, $P_{mn,y}^A=\left(\frac{2n-N-1}{2}\right)b$, $n= 1,2,\cdots,N$, and $P_{mn,z}^A=0$. For the $mn$th element, the geometry is shown in the lower right corner of Fig.~\ref{system_model}. The parameters $\mathbf{r}_{mn}^{i}$, $\mathbf{r}_{mn}^{r}$, and $\mathbf{r}_{d}$ denote the vector from Tx to the $mn$th element, the vector from the $mn$th element to Rx, and the vector from Tx to Rx, respectively. $\theta_{mn}^i$ and $\phi_{mn}^i$ represent the elevation and azimuth angles of incident wave, respectively. $\theta_{mn}^r$ and $\phi_{mn}^r$ represent the elevation and azimuth angles of reflected wave, respectively. $G_t\left(\theta_{mn}^{T}\right) $ and $G_r\left(\theta_{mn}^{R}\right) $ are the normalized transmit and receive antenna gains, respectively, where $\theta_{mn}^{T}$ and $\theta_{mn}^{R}$ are the elevation angles from the facing direction of antenna to a certain transmit and receive direction, respectively. $\theta_d^T$ and $\theta_d^R$ are the elevation angles from the facing direction of Tx antenna to a certain transmit direction and the elevation angle from the facing direction of Rx antenna to a certain receive direction, respectively. {In the following part, we assume that the main lobe of Tx directional antennas faces towards the coordinate origin (i.e., the center of the RIS) as the location of RIS is generally known by the BS. Since the Rx generally cannot acquire the position of the RIS and the BS, we assume that the Rx does not have antenna gain for brevity. Additionally, as the proposed scheme depends on the relative positions, the polarization does not affect the design of scheme, so we assume that the polarization of the transmitter and receiver are always properly matched, even after the transmitted signal is reflected by the RIS, referring to \cite{RIS_Path_Loss}. Furthermore, we do not consider the coupling effect\footnote{{In practical implementations, reflected coefficients of adjacent elements are mutually coupled, thus leading to a performance loss. Investigating this loss is beyond the scope of this work.}} between the adjacent element, which is beyond the scope of our work.}

\section{CSI-free Phase Configuration Scheme}
{In this section, we first introduce the property of Fresnel zone, with which we reveal the potential relation between the property of Fresnel zone and geometry. Then, we propose the CSI-free TPOSJ scheme and design the proper frame structure. Also, the feasibility of the proposed scheme is analyzed.}
\begin{figure*}[t]
    \centerline{\includegraphics[width=14cm]{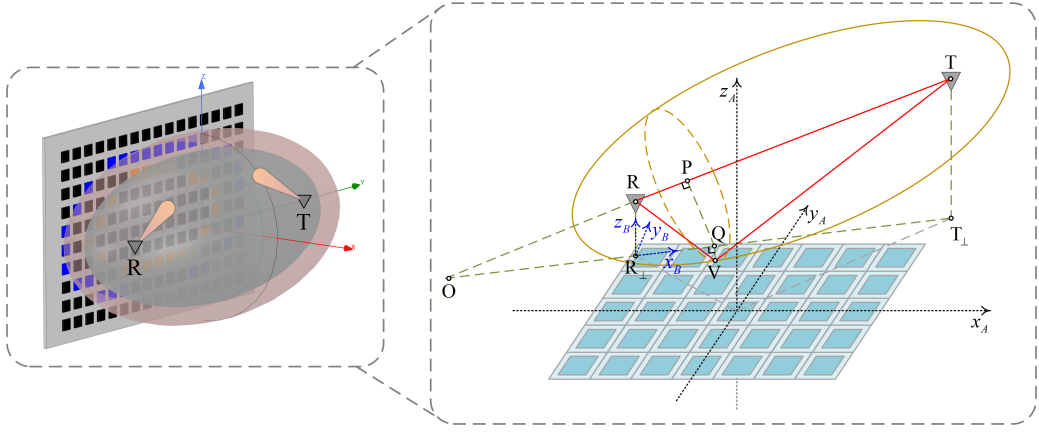}}
        \caption{{The 3D schematic diagram of the Fresnel zone and the geometric relations.}}
        \label{fresnel_geo}
        \vspace{-10pt}
\end{figure*}
\subsection{Combining Fresnel Zone with Geometry}
To leverage the position information, we present the definition and property of the Fresnel zone first. The Fresnel zone, which is based on the Huygens-Fresnel principle, describes the interference phenomenon between the reflected signal and the direct signal. The boundary of the $i$th Fresnel zone is in the shape of ellipsoid defined as the paths with $i\pi$ propagation phase shifts concerning the shortest line of sight (LoS) path, where the positions of Tx and Rx are the focal points of such ellipsoids\cite{RISFresnel}. According to the property of Fresnel zone, the reflected signal changes in phase by $\pi$ from even-numbered Fresnel zones resulting in destructive interference. Conversely, the reflected signal changes in phase by $2\pi$ from odd-numbered Fresnel zones resulting in constructive interference. As shown in the left part of Fig.~\ref{fresnel_geo}, the RIS is located in the Fresnel zone across several Fresnel zone boundaries. The minimum and maximum indexes of the Fresnel zone intersecting with the RIS are denoted by $i_l$ and $i_u$, respectively. 

Figure~\ref{fresnel_geo} also displays the geometry of Fresnel zone boundary among Tx, Rx, and RIS, where T and R represent the transmitter and receiver, respectively. As shown in the right part of Fig.~\ref{fresnel_geo}, the dotted circle represents the sectional boundary of the Fresnel zone. We denote by $\rm R_{\bot}$ and $\rm T_{\bot}$ the projection points of R and T on the $xy$-plane, respectively. $\rm P$ is the arbitrary point in straight line ${\rm RT}$, which satisfies $\left\lvert {\rm RP}\right\rvert=d$, $\left\lvert {\rm RP}\right\rvert=d_1$, and $\left\lvert {\rm PT}\right\rvert=d_2$. Therefore, ${\rm R_{\bot}T_{\bot}}$ is the projection line of ${\rm RT}$ in the $xy$-plane. The point $\rm Q$ satisfies ${\rm PQ}\bot{\rm RT}$ and ${\rm QV}\bot{\rm R_{\bot}T_{\bot}}$. The point $\rm V$ is the intersection between Fresnel boundary and the $xy$ plane. The length of ${\rm PV}$, denoted by $\rho_i$, is called the Fresnel radius of the $i$th Fresnel zone which relates to $d_1$ and $d_2$. The point $\rm O$ is the intersection between the extended lines of ${\rm R_{\bot}T_{\bot}}$ and ${\rm RT}$.

The red lines in Fig.~\ref{fresnel_geo} are the direct link and reflected link. The length of reflected link can be expressed as
\begin{align}
    \left\lvert {\rm RV}\right\rvert+\left\lvert {\rm VT}\right\rvert=&\sqrt{d_1^2+\rho_i^2}+ \sqrt{d_2^2+\rho_i^2} \\
    &\approx d_1+d_2+\frac{1}{2} \rho_i^2 \left(\frac{1}{d_1}+\frac{1}{d_2}\right),
\end{align}
where the approximation is based on Taylor expansion with $d_1,d_2\gg\lambda$. Thus, the distance difference between the reflected link and direct link can be expressed as
\begin{equation}
 \begin{aligned}
    \left\lvert {\rm RV}\right\rvert+\left\lvert {\rm VT}\right\rvert- \left\lvert {\rm RT}\right\rvert=\frac{1}{2} \rho_i^2 \left(\frac{1}{d_1}+\frac{1}{d_2}\right)=\frac{i\lambda}{2}.
 \end{aligned}   
\end{equation}
Therefore, $\rho_i$ can be expressed as
\begin{align}
    \rho_i \approx  \sqrt{\frac{i\lambda d_1d_2}{d_1+d_2}}.
\end{align}

Also, as shown in Fig.~\ref{fresnel_geo}, the projection point of $\rm R$, denoted by $\rm R_\bot$, is chosen as the origin of the $B$ coordinate system for convenient calculations. The $z$-axis of the $B$ coordinate system is perpendicular to the RIS surface and the $x$-axis of the $B$ coordinate system coincides with $\rm R_\bot T_\bot$. The coordinate points in different coordinate systems can be associated by coordinate transformation. The coordinate transformation matrix, denoted by $\boldsymbol{{\rm R}}\left(\chi\right)$, can be expressed as
\begin{align}
    \boldsymbol{\rm R}\left(\chi\right)=
    \begin{bmatrix}
        \cos \chi &\sin \chi & 0 \\
        -\sin \chi &\cos \chi& 0\\
        0& 0 & 1 \label{rotation}
    \end{bmatrix},
\end{align}
where $\chi$ is the angle between the x-axes of the $A$ and $B$ coordinate systems with $\chi$ being $\arctan {\left(\left({ y_t^A-y_r^A }\right) /\left({ x_t^A-x_r^A }\right) \right) }$. Thus, the coordinates of Tx, Rx, and RIS elements in the $B$ coordinate system, denoted by ${\rm R}^B(x_r^B,y_r^B,z_r^B)$, ${\rm T}^B(x_t^B,y_t^B,z_t^B)$, and ${\rm P}_{mn}^B(P_{mn,x}^B, P_{mn,y}^B, P_{mn,z}^B)$, can be written as 
\begin{align}
    \begin{cases}
       &{\rm R}^B(x_r^B,y_r^B,z_r^B) ={\rm R} ^A \left(\boldsymbol{\rm R} \left(\chi\right)\right)^T  -\left(x_r^A, y_r^A, 0\right) ;\\
       &{\rm T}^B(x_t^B,y_t^B,z_t^B)={\rm T} ^A \left(\boldsymbol{\rm R} \left(\chi\right)\right)^T -\left(x_r^A, y_r^A, 0\right)  ;\\
       &{\rm P}_{mn}^B(P_{mn,x}^B, P_{mn,y}^B, P_{mn,z}^B)=\\
        &~~~~~~~~~~~~~~~~~~~~~{\rm P} _{mn}^A \left(\boldsymbol{\rm R} \left(\chi\right)\right)^T -\left(x_r^A, y_r^A, 0\right),
        \label{trans}
    \end{cases}
\end{align}
respectively, where $\left(\cdot\right)^T $ denotes the transpose.

The key to configuring the ON/OFF states of RIS elements is the coordinate of point ${\rm V}$, denoted by ${\rm V}\left(V^B_x,V^B_y,V^B_z\right)$, which is the intersection point between RIS and the Fresnel zone boundary. Note that we have $V^B_x=\left\lvert \rm R_\bot Q\right\rvert $, $V^B_y=\left\lvert \rm QV\right\rvert$, and $V^B_z=0$ in the $B$ coordinate system. Since $\rm QV$ is perpendicular to $\rm PQ$, we can derive $\left\lvert\rm  QV\right\rvert $ based on triangular relation. For the lengths of $\rm PQ$ and $\rm R_\bot Q$, we have the following theorem:
{\begin{theorem}
The lengths of $\rm PQ$ and $\rm R_\bot Q$ are
\begin{align}
 \left\lvert {\rm PQ}\right\rvert=\frac{z_r^B d_2+z_t^B d_1}{\left\lvert x_t^B-x_r^B\right\rvert }
\end{align}
and
\begin{align}
    \left\lvert  {\rm R_\bot Q}\right\rvert=\frac{\left(z_r^Bd_2+z_t^Bd_1\right)d-z_r^B\left(x_t^B-x_r^B\right)^2}{\left\lvert z_t^B-z_r^B\right\rvert  \left\lvert x_t^B-x_r^B\right\rvert },
\end{align}
respectively.
\end{theorem}

\textit{Proof:} See Appendix A. $\hfill\blacksquare$ 

Thus, the length of QV can be derived based on triangular relation as 
\begin{align}
    \left\lvert \rm QV\right\rvert =\sqrt{\frac{i\lambda d_1d_2}{d_1+d_2}-{\left(\frac{z_r^B d_2+z_t^B d_1}{\left\lvert x_t^B-x_r^B\right\rvert }\right) ^2}}.
\end{align}
The ellipsoid with a fixed Fresnel radius will intersect with the RIS plane at two points given by
\begin{equation}
   \begin{aligned}
   {\rm V}=
    \begin{pmatrix}
        \frac{\left(z_r^Bd_2+z_t^Bd_1\right)d-z_r^B\left(x_t^B-x_r^B\right)^2}{\left\lvert z_t^B-z_r^B\right\rvert  \left\lvert x_t^B-x_r^B\right\rvert }\\ 
        \sqrt{\frac{i\lambda d_1d_2}{d_1+d_2}-{\left(\frac{z_r^B d_2+z_t^B d_1}{\left\lvert x_t^B-x_r^B\right\rvert }\right) ^2}}\\ 
        0 
    \end{pmatrix}^\top \label{co1}
\end{aligned} 
\end{equation}
and
\begin{equation}
   \begin{aligned}
    {\rm V}=
     \begin{pmatrix}
         \frac{\left(z_r^Bd_2+z_t^Bd_1\right)d-z_r^B\left(x_t^B-x_r^B\right)^2}{\left\lvert z_t^B-z_r^B\right\rvert  \left\lvert x_t^B-x_r^B\right\rvert }\\ 
         -\sqrt{\frac{i\lambda d_1d_2}{d_1+d_2}-{\left(\frac{z_r^B d_2+z_t^B d_1}{\left\lvert x_t^B-x_r^B\right\rvert }\right) ^2}}\\ 
         0 
     \end{pmatrix}^\top, \label{co2}
 \end{aligned} 
\end{equation}
respectively.}

\subsection{TPOSJ Scheme}
According to the property of the Fresnel zone, the elements located at the odd and even Fresnel zone should be designed as OFF and ON states, respectively. {In addition, for these elements located at the center between two adjacent Fresnel zones, we should control them as absorption mode, since the reflected signals from these elements are indistinct. When position error occurs, these elements are more likely to provide the opposite effect, which will attenuate the received power. The judgement function of RIS element at the $m$th row and the $n$th column at the $i$th Fresnel zone, denoted by $\varOmega_{m,n}^i$, is given by
\begin{equation}
   \begin{aligned}
    \varOmega_{m,n}^i=
    \begin{cases}
        \text{if}~~i = {\rm odd} &\begin{cases}
            &{\mit \Gamma}_{r},~~\text{if} ~~~{\left\lvert{\rm P}_{mn}^B-{\rm V} \right\rvert} \leqslant \xi;  \\
            &{\mit \Gamma}_{a},~~\text{if} ~~~{\left\lvert{\rm P}_{mn}^B-{\rm V} \right\rvert} > \xi ;
            \end{cases}\\[0.6cm]
        \text{if}~~i = {\rm even}   &\begin{cases}
            &-{\mit \Gamma}_{r},~~\text{if} ~~~{\left\lvert{\rm P}_{mn}^B-{\rm V} \right\rvert} \leqslant \xi ; \\
            &{\mit \Gamma}_{a},~~\text{if} ~~~{\left\lvert{\rm P}_{mn}^B-{\rm V} \right\rvert} > \xi ,
            \end{cases}
    \end{cases}
    \label{judge}
\end{aligned} 
\end{equation}
where $\xi$ represents the threshold.} The reflected signals from the ON state elements are more comparable to the LoS path as $\xi$ decreases. However, this will result in fewer elements in reflection mode. Since the reflected signal within the phase difference from $-\pi$/2 to $\pi$/2 can enhance the received power based on the interference condition, the general value of maximum allowable phase difference is $\pi$/2. The threshold $\xi$ can be expressed as follows:
\begin{equation}
        \begin{aligned}
            \xi=&\max_{d_1,d_2,i} \sqrt{ \frac{(i+\tau )\lambda d_1d_2}{d_1+d_2}-\left(\frac{z_r^B d_2+z_t^B d_1}{x_t^B-x_r^B}\right) ^2}\\
            &~~~~~~~~~~~~~~~~~-\sqrt{ \frac{i\lambda d_1d_2}{d_1+d_2}-\left(\frac{z_r^B d_2+z_t^B d_1}{x_t^B-x_r^B}\right) ^2} \\
            &{\rm s.t.}~~~~~~~~ \frac{i\lambda d_1d_2}{d_1+d_2}-\left(\frac{z_r^B d_2+z_t^B d_1}{x_t^B-x_r^B}\right) ^2>0,
        \end{aligned}   
        \label{threshold} 
\end{equation}
where $\tau \in(0,~0.5]$. {With a smaller value of $\tau$, there will be more elements in absorption mode and less elements in reflection mode. Nevertheless, the reflected signals from these elements in reflection mode have strong constructive interference with a good robustness performance. In general, based on interference condition, $\tau$ can be set as 0.5. Then, the received power can achieve its maximum value and no elements are in absorption mode. However, considering the position error, the value of $\tau$ should be properly choosed to satisfy the desirable robustness performance between achievable rate and robustness to position error. }

Thus, the RIS configuration matrix for the $i$th Fresnel zone, denoted by $\boldsymbol{\Omega}^i$, can be expressed as
\begin{equation}
  \begin{aligned}
    \boldsymbol{\Omega}^i =
    \begin{bmatrix}
        \varOmega_{1,1}^i&\varOmega_{1,2}^i&\cdots&\varOmega_{1,N}^i\\
        \varOmega_{2,1}^i&\varOmega_{2,2}^i&\cdots&\varOmega_{2,N}^i\\
        \vdots&\vdots&\ddots&\vdots\\
        \varOmega_{M,1}^i&\varOmega_{M,2}^i&\cdots&\varOmega_{M,N}^i
    \end{bmatrix}. \label{RIS}
\end{aligned}  
\end{equation}
We define the function $\Upsilon\left(\rm \boldsymbol{A}\right)$, which substitutes each element of matrix $\rm \boldsymbol{A}$ into $\varepsilon \left(t-{\mit \Gamma}_{r}\right)-\varepsilon \left(-{\mit \Gamma}_{r}-t\right)$ and outputs the same size matrix where the function $\varepsilon\left(\cdot\right)$ representing the step function. Then, the final RIS configuration matrix, denoted by $\boldsymbol{\Omega}$, can be expressed as
\begin{equation}
  \begin{aligned} 
    \boldsymbol{\Omega}=
        \Upsilon  \left(\sum _{j\in \left[i_l, i_u\right] } \boldsymbol{\Omega}^j\right)
        =
        \begin{bmatrix}
        \varOmega_{1,1} &\varOmega_{1,2} &\cdots&\varOmega_{1,N} \\
        \varOmega_{2,1} &\varOmega_{2,2} &\cdots&\varOmega_{2,N} \\
        \vdots&\vdots&\ddots&\vdots\\
        \varOmega_{M,1} &\varOmega_{M,2} &\cdots&\varOmega_{M,N} 
    \end{bmatrix},
    \label{sum}
\end{aligned}  
\end{equation}
where the function $\sum _{j\in \left[i_l, i_u\right] }$ denotes the continuous sum among elements in the same location of different matrices from the indexes $i_l$ to $i_u$ and $\varOmega_{m,n}$ is the final phase state of the $mn$th element with $m= 1,\cdots,M$ and $n= 1,\cdots,N$. 

Based on Eqs.~\eqref{co1} and \eqref{co2}, we can conclude that the interval of adjacent Fresnel zone decreases as the index $i$ increases. Thus, there are more than one Fresnel zones intersecting with the same RIS element when the value of the element in $\sum _{j\in \left[i_l, i_u\right] } \boldsymbol{\Omega}^j$ is greater than ${\mit \Gamma}_{r}$ or less than $-{\mit \Gamma}_{r}$, revealing that the indexes of these Fresnel zones are large. In this case, the impact of the mentioned element on received power is negligible. First, the function of RIS is insufficient to permit such delicate modification of reflected signal as we cannot separately manipulate the reflected signal from multiple Fresnel bounds by single element. Second, the ON/OFF states of these RIS elements have a limited impact on the received power since the power mainly concentrates on the first few Fresnel zones\cite{RISFresnel}.

{Thus, the only remaining issue is searching the index ranges $i_l$ and $i_u$ to derive the selective indexes of the Fresnel zone $j$ which intersect with RIS array. The lower bound $i_l$ can be derived based on the coordinates of Tx and Rx as follows:
\begin{equation}
   \begin{aligned}
   &~~i_{l}=\min i\\
   {\rm s.t.}~~&\exists ~\left\lvert {\rm V}-{\rm P}_{mn}^B\right\rvert ^2< \xi. \label{lower} 
\end{aligned} 
\end{equation}
The upper bound of the index $i_u$, which relates to the coordinates of Tx and Rx as well as the RIS physical size, can be expressed by
\begin{equation}
\begin{aligned}
    ~~i_{u}=&\max i\\
    {\rm s.t.}~~\exists~\left\lvert {\rm V}-{\rm P}_{mn}^B\right\rvert ^2&< \xi;~~~{\rm P}_{mn}^A \in \mathcal{A}, \label{upper} 
\end{aligned}    
\end{equation}}
where the set $\mathcal{A}$ contains the coordinates of RIS elements located at the outermost boundary and the relation between ${\rm P}_{mn}^B$ and ${\rm P}_{mn}^A$ can be referred in Eq.~\eqref{trans}. The set $\mathcal{A}$ can be expressed as $\left\{\left({a\left(1-M\right)}/{2},{b\left(1-N\right)}/{2}, 0\right)\right.$, $\left({a\left(1-M\right)}/{2},\! {b\left(N-1\right)}/{2},\! 0\right)$,$ \left({a\left(M-1\right)}/{2},\! {b\left(1-N\right)}/{2}\right.,\! \left.0\right)$, $\left. \left({a\left(M-1\right)}/{2}, {b\left(N-1\right)}/{2}, 0 \right)\right\}$, which consists of four RIS elements located at the outermost corner, according to the fact that the maximum Fresnel zone always intersects with one of the outermost RIS elements. {With the bounds of indexes, the procedure of the phase configuration for SISO case can be summarized as follows: 

{\textbf{Procedure:~}}{\textbf{1)}} derive the lower bound $i_{l}$ and upper bound $i_{u}$ according to Eqs.~\eqref{lower} and \eqref{upper}; {\textbf{2)}} calculate the ON/OFF states of RIS elements located at the $j$th Fresnel zone in terms of Eqs.~\eqref{rotation}, \eqref{co1}, \eqref{co2}, \eqref{judge}, and \eqref{RIS} to derive $\boldsymbol{\Omega}^j $ with $j\in \left[i_l, i_u\right]$; {\textbf{3)}} derive the ultimate configuration matrix $\boldsymbol{\Omega}$ with Eq.~\eqref{sum} corresponding to the selected pair.}

{\subsection{Extension to MISO and Multi-RIS case}
In this subsection, we show that the proposed TPOSJ scheme can be easily extended to MISO and multiple RISs case.

Assume that the Tx is equipped with $B$ antennas, surrounding by $C$ RISs, while other parameters hold the same. With the known positions of RISs and Tx's multiple antennas, the RIS configuration corresponding to each pair (one of the Tx antenna and one of the RIS) can be derived by performing the above \textbf{Procedure} for each pair (SISO case). The configuration matrix for the pair comprising the $b$th antenna of Tx and the $c$th RIS, denoted by $\boldsymbol{\Omega}_b^c$ with $c\in\left[1, C\right]$ and $b\in\left[1,B\right]$, can be derived by substituting their positions into the \textbf{Procedure}. Then, the final configuration matrix for the $c$th RIS, denoted by $\boldsymbol{\Omega}^c$, can be expressed as follows:
\begin{equation}
    \begin{aligned}
        \boldsymbol{\Omega}^c=
    \Upsilon  \left(\sum _{b\in \left[1, B\right] } \boldsymbol{\Omega}_b^c\right),
    \end{aligned}
    \label{final_RIS}
\end{equation}
where $\Upsilon\left(\cdot\right)$ is defined in Eq.~\eqref{sum}. Following the same procedure, the RIS configuration matrices for all RISs can be derived. For the considered MISO and multi-RIS case, the above \textbf{Procedure} should be carried out for $BC$ times in total. Based on the above results, the processes of the TPOSJ scheme are shown in \textbf{Algorithm 1} in detail.

\vspace*{0.4cm}
\begin{algorithm}[t]
    \caption{The TPOSJ algorithm} 
    \hspace*{0.02in} {\bf Initialization:}
    $k=1$
    \begin{algorithmic}[1]
    \WHILE{for the $k$th time slot, $k\in[1,K]$}
    \STATE $c=1$;
    \WHILE{for the $c$th RIS, $c\in[1,C]$}
    \STATE {$b=1$};
    \WHILE{for the $b$th Tx antenna, $b\in[1,B]$}
    \STATE Calculate the lower bound $i_{l}$ and upper bound $i_{u}$ by Eqs.~\eqref{lower} and \eqref{upper}, and let $j=i_{l}$;
    \WHILE{$j\in [i_{l}, i_{u}]$}
    \STATE Calculate the ON/OFF states of RIS elements located at the $j$th Fresnel zone based on Eqs.~\eqref{rotation}, \eqref{co1}, \eqref{co2}, \eqref{judge}, and \eqref{RIS} to derive $\boldsymbol{\Omega}^j$;
    \STATE $j=j+1$;
    \ENDWHILE
    \STATE {Calculate the ultimate configuration matrix $\boldsymbol{\Omega}$ by Eq.~\eqref{sum} and let $\boldsymbol{\Omega}_b^c=\boldsymbol{\Omega}$};
    \STATE {$b=b+1$};
    \ENDWHILE
    \STATE{ Derive the final configuration matrix $\boldsymbol{\Omega}^c$ for the $c$th RIS by Eq.~\eqref{final_RIS}};
    \STATE {$c=c+1$};
    \ENDWHILE
    \STATE {Control the mode and state of the $c$th RIS elements according to the ultimate configuration matrix $\boldsymbol{\Omega}^c$};
    \STATE $k=k+1$;
    \ENDWHILE
    \end{algorithmic}
\end{algorithm}

}

{
\subsection{Feasibility Analysis}
In this subsection, we first analyse the complexity of the proposed TPOSJ scheme. Additionally, as the phase configuration is highly related to the position of Rx, the impacts of mobility and position error is analyzed to evaluate the its practicability.

{\textbf{\underline{Complexity analysis}:}~The position-driven scheme shows superiorities over the traditional CSI-driven scheme with respect to complexity in three aspects including computational complexity, estimation complexity and singling overhead. Firstly, for the computational complexity, in the phase configuration phase of the time slot for each user, the overall complexity contains two parts. Therein, to derive the lower bound $i_l$ and upper bound $i_u$, the complexity with respect to $M$ and $N$ is $\mathcal{O}\left(i_lMN\right)$ corresponding to step 6 in the \textbf{Algorithm 1}. Meanwhile, within the range of indexes from $i_l$ and $i_u$, the complexity of step 8 is $\mathcal{O}\left(\left(i_u-i_l\right) MN\right)$. Thus, the overall calculation complexity in terms of RIS elements is $\mathcal{O}\left(i_uMN\right)$. For MISO and multi-RIS case, the overall complexity is $\mathcal{O}(MN\sum_{b=1}^{b=B}\sum_{c=1}^{c=C}i_{b,u}^c)$, which increases linearly as RIS elements increase. Therein, $i_{b,u}^c$ is the upper bound of the intersected Fresnel zone for the pair of $b$th antenna and the $c$th RIS. In contrast, the complexity of optimization with the semidefinite relaxation (SDR) is $\mathcal{O}(\left(MN+1\right) ^4\sqrt{MN})$ with respect to the number of RIS. Secondly, the estimation of priori information (i.e., CSI or position) is essential to design the RIS configurations. In the position-driven frame, the traditional CSI estimation is replaced by position estimation in each time slot. The complexities of RIS-assisted user localization and RIS-associated channel estimation are $\mathcal{O}\left(MN\right)$ and $\mathcal{O}\left(M^2N^2\right)$ with respect to the number of RIS elements, respectively\cite{complexity}. Thirdly, estimating the priori information is based on pilot symbols. For RIS-assisted CSI-driven scheme in the near-field, the same orthogonal pilot set is used to estimate the channel parameters for massive RIS elements.  The frequency of pilot reuse increases as the number of RIS elements increases and the estimation performance suffers from the pilot pollution\cite{Pilot_pollution}. In contrast, in position-driven scheme, the pilot symbol is allocated to the specific UE, which is irrelevant to the RIS. In summary, according to these superiorities, our proposed scheme can significantly facilitate the timely response.}

{\textbf{\underline{The impact of mobility}:}}~To investigate the impact of mobility, we divide the mobility in low mobility (e.g., human, vehicle in urban environments) and high mobility (e.g., over 40 km/h). In long term evolution (LTE) networks and 5G new radio (NR), the slot time is 1 ms. The moving distance in one slot time is within $0.0011 ~{\rm m} \sim 0.011~{\rm m}$ with the moving speed being from $ 4~{\rm km/h}$ to $40 ~{\rm km/h}$ in typical urban environments, which is far less than the localization accuracy in 5G networks. Under this circumstance, the update rate of the RIS configuration in each time slot is feasible. Thus, in low mobility scenario, the position deviation in one slot can be regarded as position error. However, in high mobility scenario, the update rate of RIS configuration in typical time slot may be insufficient. It is noteworthy that as the mobility of Rx increases, the trajectory of motion is more predictable. The extra information of velocity and direction should be obtained at Tx by proper design, e.g., existing scheme about velocity and direction estimation, or feedback mechanism in each time slot\cite{speed_estimation}. Thus, we only need to add the velocity and direction estimation or velocity feedback procedures into the phase of RIS-assisted localization in each time slot while other phases stay the same.}

{
{\textbf{\underline{The impact of position error}:}}~Consider the practical scenario that the Tx has the accurate position of the RIS and itself, while the estimated position of Rx is inaccurate with Gaussian distribution errors\cite{Gaussian_Error}, which is denoted by ${\rm R}^B_e\left(x_e^B,y_e^B,z_e^B\right)$. The estimated position can be expressed as follows:
\begin{align}
    {\rm R}^B_e\left(x_e^B,y_e^B,z_e^B\right)={\rm R}^B(x_r^B,y_r^B,z_r^B)+e(e_x,e_y,e_z),
\end{align}
where $e(e_x,e_y,e_z)$ denotes the 3D Gaussian distribution error with $e_x$, $e_y$, and $e_z$ being the error components regarding $x$-axis, $y$-axis, and $z$-axis, respectively. It is difficult to directly analyze the impact of position error based on Eqs.~\eqref{co1} and \eqref{co2}. Thus, we leverage the ellipsoid equation of Fresnel zone to evaluate the impact of position error on intersection point ${\rm V}$ as the Fresnel zone is in the shape of ellipsoid. We denote by $\rm U$ as the midpoint between the Tx and Rx, which can be expressed by
\begin{align}
    {\rm U}=\left(U_x,U_y,U_z\right) =\left(\frac{x_e^B+x_t^B}{2},\frac{y_e^B+y_t^B}{2},\frac{z_e^B+z_t^B}{2}\right),
\end{align}
where $U_x,~U_y$ and $U_z$ are the components with respect to the coordinate. Thus, we can derive the equation of the $i$th Fresnel zone with $\nu \in \left[0,\pi\right] $ and $\kappa \in \left[0,2\pi\right]$ as follows:
\begin{align}
    \begin{pmatrix}
        x_i & y_i & z_i 
    \end{pmatrix}^T =
    {\rm U}^T+
    \boldsymbol{{\rm R}}_y\left(\beta \right) 
    \begin{pmatrix}
        l_a &0& 0\\0 &l_b& 0\\0 &0& l_c
    \end{pmatrix}
    \begin{pmatrix}
        \sin \nu   \cos \kappa\\ \sin \nu   \sin \kappa\\\cos \nu   
    \end{pmatrix}, \label{xyz}
\end{align}
where $\left(x_i ,y_i,z_i\right) $ is the coordinate of the $i$th Fresnel zone, $\boldsymbol{{\rm R}}_y$ is the rotation matrix with respect to $y$-axis, $\beta$ is the elevation angle between the line $\rm TR$ and the $xoy$-plane with $\beta=\arctan\left(\left(z_e^B-z_t^B\right)/\left(x_e^B+x_t^B\right)  \right) $, $l_a$, $l_b$, $l_c$ are the semi-major axis and two semi-minor axes, respectively. $\boldsymbol{{\rm R}}_y\left(\beta \right)$ can be expressed as follows:
\begin{align}
    \boldsymbol{{\rm R}}_y\left(\beta \right) =
    \begin{pmatrix}
        \cos \beta & 0& -\sin\beta \\
         0 & 1 & 0\\
        \sin \beta& 0 &  \cos\beta
    \end{pmatrix}. \label{Rz}
\end{align}
According to the property of the Fresnel zone, $l_a=d+i\lambda/{2}$, $l_b=i\lambda d/2$, and $l_c=i\lambda d/2$. Thus, we can derive the coordinate of equation as follows:
\begin{align}
    \begin{cases}
        x_i=\left ( d+\frac{ i\lambda}{2} \right ) \cos \beta \sin \nu   \cos \kappa-\frac{i\lambda d \sin \beta \cos \nu }{2}+\frac{x_e^B+x_t^B}{2};\\[0.1cm]
        y_i=\frac{i\lambda d \sin \nu \sin \kappa }{2} +\frac{y_e^B+y_t^B}{2};\\[0.1cm]
        z_i=\left ( d+\frac{ i\lambda}{2} \right )\sin \beta \sin \nu   \cos \kappa +\frac{i\lambda d \cos \beta \cos \nu }{2}+\frac{z_e^B+z_t^B}{2}.
        \end{cases}
 \end{align}
When $z_i=0$, we can derive the x-coordinate and y-coordinate of intersection line as 
\begin{equation}
  \begin{aligned}
    x_i&=U_x+U_z\tan\beta + \frac{l_a\sin \nu   \cos \kappa}{\cos \beta}\\
    &=\frac{x_r^B+x_t^B+\tan\beta\left(z_r^B+z_t^B\right) }{2} +\frac{l_a\sin \nu   \cos \kappa}{\cos \beta}\\
    &~~~~~~~~~~~~~~~~~~~~~~~~~~~~~~~~~~+\underbrace{\frac{e_x+e_z\tan\beta }{2}}_{\rm error}  \label{xerror}
\end{aligned}   
\end{equation}
and 
\begin{align}
    y_i=\frac{y_r^B+y_t^B}{2}+ l_b \sin \nu   \sin \kappa +\underbrace{\frac{e_y}{2}}_{\rm error}, \label{yerror}
\end{align}
respectively. Thus, the position error will have an impact on $\beta$, $x_i$, and $y_i$. {In general scenarios that the RIS is distant from Tx and Rx, the transmission distance is several orders than the position error, and $\beta$ can be approximated as $\beta \approx \arctan\left(\left(z_r^B-z_t^B\right)/\left(x_r^B-x_t^B\right) \right)$.} Based on Eqs.~\eqref{xerror} and \eqref{yerror}, both $x_i$ and $y_i$ are mainly disturbed by Gaussian position error and slightly impacted by the approximation of $\beta$, where the noise term in $x_i$ is amplified by the factor $\tan \beta$. Thus, in the deployment example, the impact of error on $x$-coordinate is more severe than the impact of error on $y$-coordinate. The impact of error on $z$-coordinate is mainly induced by the approximation of $\beta$, which is relatively minimal. 

\begin{remark}
    When the altitudes of Tx and Rx are the same, the TPOSJ scheme can achieve the best performance since $\tan \beta=0$. However, when the distance between Tx and Rx is very close and the height difference is large, the noise term regarding $x_i$ will be amplified. Thus, different localization precision about different coordinates should be adopted to prevent this condition. 
\end{remark}
}

\begin{figure}[t]
    \centerline{\includegraphics[width=6cm]{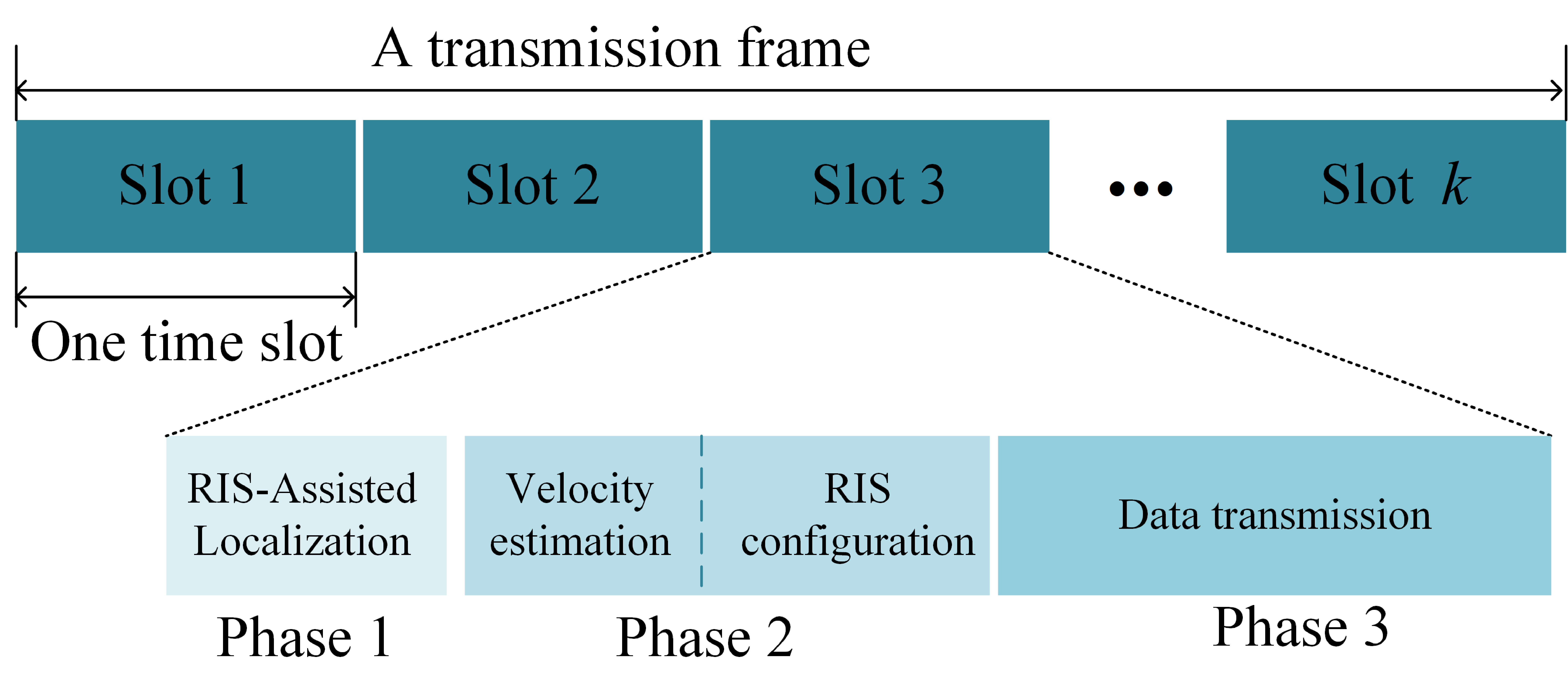}}
    \vspace{-10pt}
        \caption{{The proposed position-driven frame structure.}}
        \label{frame}
        \vspace{-10pt}
\end{figure}

\subsection{Position-Driven Frame Structure}
{The proposed location-driven scheme align well with industrial internet of things (IoT) in 3GPP Release 16 industry 4.0\cite{3GPP_16}. In specific, in future industrial IoT networks, various services including sensors, automatic car, and mobile terminals are required to support the crucial demands including timely transmissions and mobile connections. We elaborate on two superiorities according to the above two demands by integrating the recent proposals in standardization organization 3GPP. Firstly, time sensitive communications are essential for precise manipulation to prevent collisions among various automated terminals in IoT networks. To support the timely transmissions, according to the \textbf{\underline{Complexity analysis}}, the TPOSJ scheme has low complexity including computational complexity, estimation complexity, and singling overhead compared with traditional CSI-driven scheme, thus facilitating timely responses for time sensitive demands and reducing the hardware cost for processor. Secondly, the mobility support is another distinct feature of intelligent industrial networks which offers service continuity to automated moving terminals. Due to outdated CSI, UE with medium or high mobility experiences significant performance loss. To support such demand, the potential CSI reporting enhancements are explored in 3GPP Release 18, which necessitates frequent updates of CSI estimation. In contrast, the UE's mobility can be naturally captured in location-driven frame by trajectory prediction techniques and feedback mechanisms. Especially for future industrial IoT networks, the motion and velocity of automated terminals are always sensed and monitored. Then, the configurations can be predesigned and updated to further boost the response speed and support the demands for mobility. Additionally, the positioning metrices are progressively improved from 3GPP Release 16 to Release 18. Especially with the extra degrees of freedom provided by the RIS, localization for UE can be enhanced in cellular network by considering the presence of RIS according to ETSI GR RIS 002 V1.1.1 (2023-08), which well supports the prior information required by position-driven scheme.} On this basis, we design the position-driven frame structure, as shown in Fig.~\ref{frame}, where one frame with $k$ time slots can be assigned to different users. Each time slot consists of three parts. {In phase 1, the Tx performs RIS-assisted localization to derive priori position information\cite{Location-Awareness,siso_localization3}. In phase 2, with the improved positioning and velocity sensing parameters, the Tx implements the proposed TPOSJ scheme to design the RIS configuration matrix and controls it by a RIS controller.} The data transmission is enabled with the configured phase shifts of RIS in phase 3. The same process is repeated in different time slots.

\vspace{-5pt}
\section{Huygens-Fresnel-based Power Flow}
In this section, as the channel gains should be known previously in order to perform power allocation and transmission rate control, we derive the power flow of individual element and RIS array based on Huygens-Fresnel principle and Helmholtz-Kirchhoff integral theorem, which gives more general expressions for the power scaling law of RIS.
\begin{figure}[t]
    \centering
    \begin{minipage}[t]{0.48\textwidth}
    \centering
    \includegraphics[width=5cm]{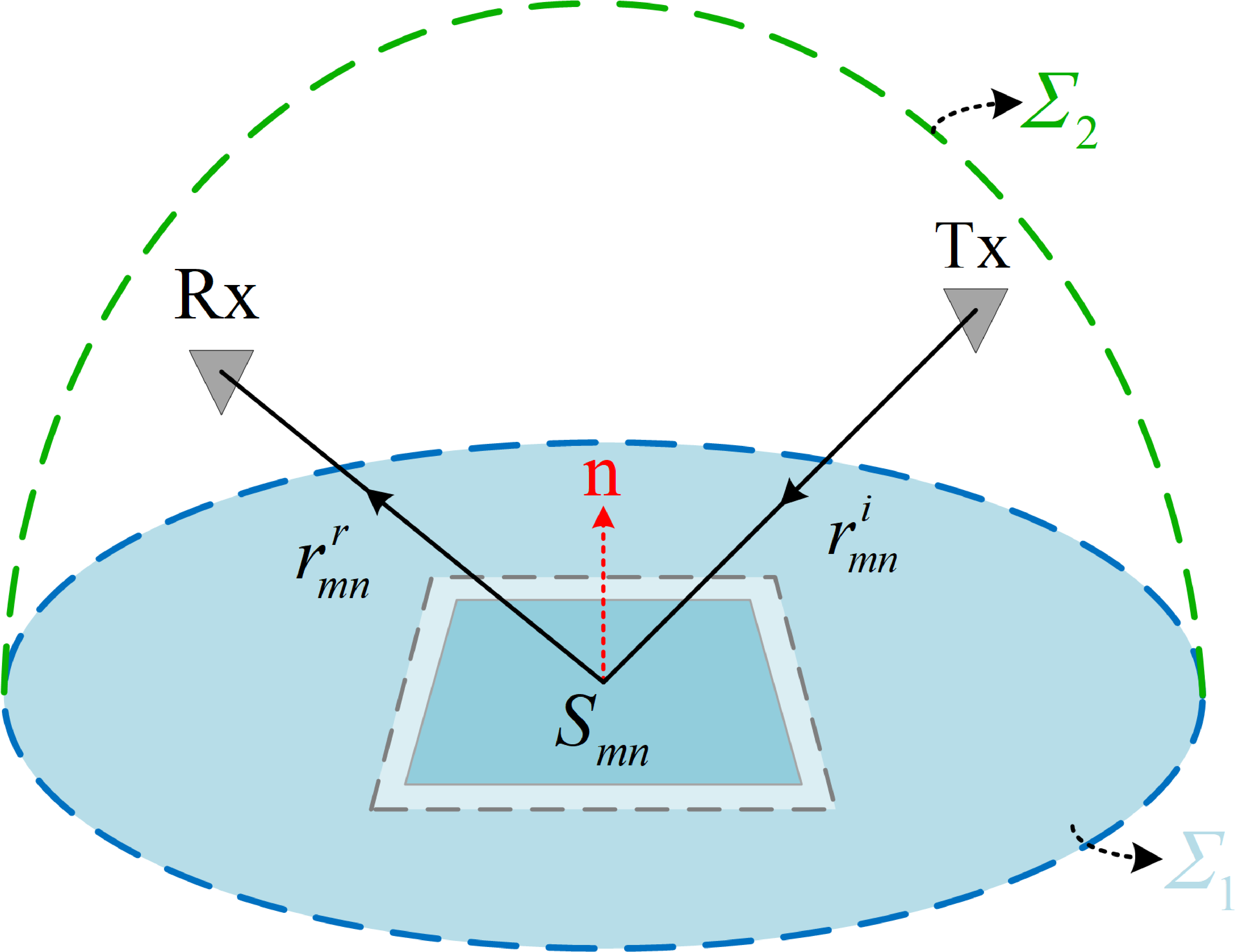}
    \caption{The closed boundary of Rx, Tx, and individual RIS element.}
    \label{Huygens}
    \end{minipage}
    \begin{minipage}[t]{0.48\textwidth}
    \centering
    \includegraphics[width=5cm]{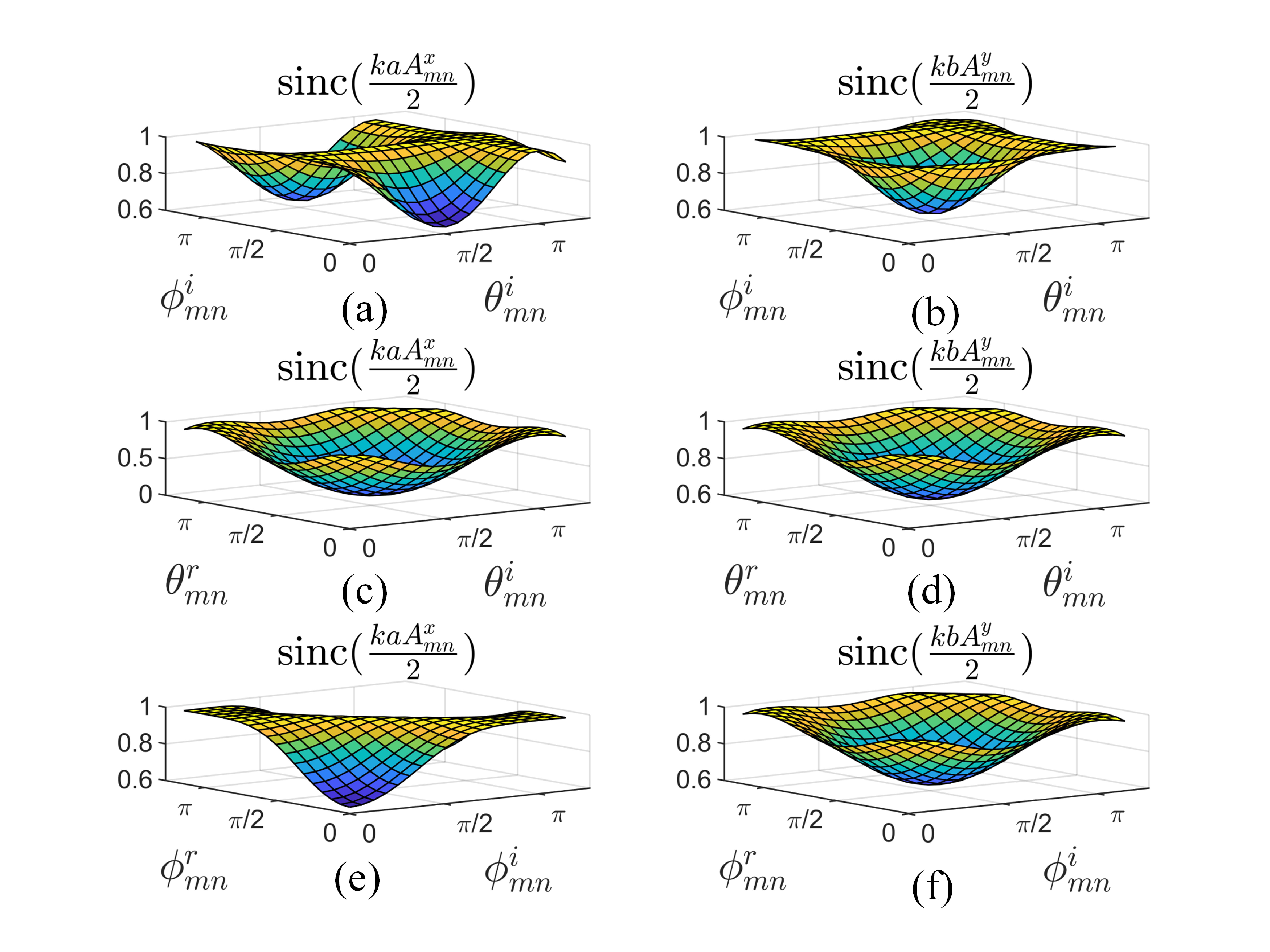}
    \caption{The impact of azimuth and elevation angle on the sinc function. (a), (b) The impact of $\theta^i_{mn}$ and $\phi^i_{mn}$ when $\theta^i_{mn}=\phi^i_{mn}=0$. (c), (d) The impact of $\theta^i_{mn}$ and $\theta^r_{mn}$ when $\phi^i_{mn}=\phi^r_{mn}=\pi/6$. (e), (f) The impact of $\phi^i_{mn}$ and $\phi^r_{mn}$ when $\theta^i_{mn}=\theta^r_{mn}=\pi/6$.}
    \label{sinc}
    \end{minipage}
    \vspace{-10pt}
\end{figure}

{The RIS element can be regarded as an individual entity by ignoring the coupling effect. Thus, to begin with, we need to obtain the reflected electric field induced by individual RIS element. The Huygens-Fresnel principle, which is suitable for analyzing scalar wave propagation problems, can be applied to derive the electric field as the vector wave can be separated into scalar waves in different directions\cite{Huygens-Fresnel}. However, the authors in \cite{Huygens-Fresnel} only consider the 2-D plane. Thus, in this section, we leverage the Helmholtz-Kirchhoff integral theorem, which gives a more rigorous mathematical expression based on wave equation for Huygens-Fresnel principle, to derive the power flow of RIS. }According to the Helmholtz-Kirchhoff integral theorem, the reflected electric field at the observation point can be expressed in terms of the electric field and its derivative on any closed surface surrounding the observation point. Fig.~\ref{Huygens} depicts an example of a closed boundary that contains the transmitter, receiver, and RIS element, where $S_{mn}$ denotes the surface of the $mn$th element. {$\varSigma_1$ is a disc with an infinite radius including $S_{mn}$, and $\varSigma_2$ is a hemisphere with an infinite radius above $S_{mn}$.} $\mathbf{n}$ is the normal vector of closed surface. Thus, we have the following theorem.

\begin{theorem}
The reflected electric field from $S_{mn}$, denoted by ${E}_{mn}^r$, can be expressed as follows:
\begin{equation}
    \begin{aligned}
    &{{E}}_{mn}^r=\frac{jk{\mit \Gamma} _{mn}A_{mn}}{4\pi} \times\\
    & \iint_{S_{mn}} \left\{G\left(\mathbf{r}_{mn}^{i}\right) G\left(\mathbf{r}_{mn}^{r}\right) \left[\cos\left( \mathbf{r}_{mn}^{r}, \mathbf{n}\right) -\cos\left( \mathbf{r}_{mn}^{i}, \mathbf{n}\right)\right] \right\} d\sigma, \label{electric_field}
\end{aligned}
\end{equation}
where $A_{mn}$ is the electric field intensity, ${\mit \Gamma} _{mn}$ is the reflected coefficient of the $mn$th element, $k=\frac{2\pi}{\lambda}$ is the wave number, $\lambda$ is the wavelength, $G\left(\mathbf{r}\right)=\frac{e^{-jk{r}} }{{r}}$ is the Green's function with $r=\left\lvert \mathbf{r}\right\rvert$ representing the magnitude of the vector $\mathbf{r}$, and the function $\cos\left(\mathbf{r},\mathbf{n}\right)$ denotes the cosine of included angle between vectors $\mathbf{r}$ and $\mathbf{n}$.
\end{theorem}

\textit{Proof:} See Appendix B. $\hfill\blacksquare$ 

The electric field intensity $A_{mn}$ satisfies the following equation according to Friis transmission equation:
\begin{align}
    \frac{A_{mn}^2}{2\eta_0}=\frac{P_t G_t\left(\theta_{mn}^{T}\right) }{4\pi},
\end{align}
where $P_t$ is the transmit power. The distance from Tx to the $mn$th element and the distance from the $mn$th element to Rx, denoted by ${r}_{mn}^{i}$ and ${r}_{mn}^{r}$, respectively, can be expressed as
\begin{equation}
\begin{aligned}
{r}_{mn}^{i}&= \sqrt{\left(x_t^A-x\right)^2+\left(y_t^A-y\right)^2+{z_t^A}^2} \\ 
&\approx {r}_{0,mn}^{i}-\sin\theta_{mn}^i \cos \phi_{mn}^ix- \sin\theta_{mn}^i \sin \phi_{mn}^i y \label{app1}
\end{aligned} 
\end{equation}
and 
\begin{equation}
\begin{aligned}
    {r}_{mn}^{r}&= \sqrt{\left(x_r^A-x\right)^2+\left(y_r^A-y\right)^2+{z_r^A}^2} \\
    &\approx {r}_{0,mn}^{r}-\sin\theta_{mn}^r \cos \phi_{mn}^rx- \sin\theta_{mn}^r \sin \phi_{mn}^r y, \label{app2}
\end{aligned}    
\end{equation}
respectively, where $x$ and $y$ are the x-coordinate and y-coordinate of an arbitrary point at the surface of the $mn$th element satisfying $\left(x,y,0\right)\in S_{mn}$. The parameters ${r}_{0,mn}^{i}$ and ${r}_{0,mn}^{r}$ denote the distance from Tx to the central point of the $mn$th element and the distance from the central point of the $mn$th element to Rx, respectively, which can be calculated by ${r}_{0,mn}^{i}=\text{dis}\left({\rm T}^A, {\rm P}_{mn}^A\right)$ and ${r}_{0,mn}^{r}=\text{dis}\left({\rm R}^A, {\rm P}_{mn}^A\right)$, respectively, with the function $\text{dis}\left(\rm P, Q\right)$ representing the distance between coordinate points ${\rm P}$ and ${\rm Q}$. The approximations in Eqs.~\eqref{app1} and \eqref{app2} are obtained by far-field condition of individual element. Thus, the electric field in Eq.~\eqref{electric_field} can be expressed as Eq.~\ref{dedede}, where the function sinc$\left(x\right)=\left(\sin x\right) /{x} $ denotes the sinc function of the variable $x$, $A_{mn}^x=\sin\theta_{mn}^i \cos \phi_{mn}^i+\sin\theta_{mn}^r \cos \phi_{mn}^r$, and $A^y_{mn}=\sin\theta_{mn}^i \sin \phi_{mn}^i+\sin\theta_{mn}^r \sin \phi_{mn}^r$. The factor $K^{\theta}_{mn}=\cos\theta_{mn}^r+\cos\theta_{mn}^i$ is the leaning factor in Fresnel-Kirchhoff diffraction formula. {In addition, the approximation $\left(a\right)$ is derived by far-field condition of individual element in Eqs.~\eqref{app1} and \eqref{app2}, and the equation $\left(b\right)$ holds as $\int_{-a/2}^{a/2} e^{-jkx}dx=a {\rm sinc}\left(\frac{ka}{2}\right)$ holds.}

\addtocounter{equation}{1}
\begin{figure*}[!bt]
    { 
    \begin{equation}
        \begin{aligned}
       {{E}}_{mn}^r&=\int_{-a/2}^{a/2}\int_{-b/2}^{b/2}\frac{{\mit \Gamma} _{mn}A_{mn}e^{-jk\left({r}_{mn}^{r}+{r}_{mn}^{i}\right)}}{2j\lambda {r}_{0,mn}^{r} {r}_{0,mn}^{i}} \left(\cos\theta_{mn}^r+\cos\theta_{mn}^i\right)dx dy \\  
        &\overset{\left(a\right) }{\approx} \frac{{\mit \Gamma} _{mn}A_{mn}\left(\cos\theta_{mn}^r+\cos\theta_{mn}^i\right)}{2j\lambda {r}_{0,mn}^{r} {r}_{0,mn}^{i}} \times \\
        &\int_{-a/2}^{a/2}\int_{-b/2}^{b/2} e^{-ik\left({r}_{0,mn}^{i}-\sin\theta_{mn}^i \cos \phi_{mn}^ix- \sin\theta_{mn}^i \sin \phi_{mn}^i y+{r}_{0,mn}^{r}-\sin\theta_{mn}^r \cos \phi_{mn}^rx- \sin\theta_{mn}^r \sin \phi_{mn}^r y\right)}dxdy\\
        &=\frac{{\mit \Gamma} _{mn}A_{mn}\left(\cos\theta_{mn}^r+\cos\theta_{mn}^i\right)e^{-jk\left({r}_{0,mn}^{i}+{r}_{0,mn}^{r}\right) }}{2j\lambda {r}_{0,mn}^{r} {r}_{0,mn}^{i}} \times\\ 
         &\int_{-a/2}^{a/2}e^{jk\left(\sin\theta_{mn}^i \cos \phi_{mn}^i+\sin\theta_{mn}^r \cos \phi_{mn}^r\right)x}dx \int_{-b/2}^{b/2}e^{jk\left(sin\theta_{mn}^i \sin \phi_{mn}^i+\sin\theta_{mn}^r \sin \phi_{mn}^r\right)y }dy\\ 
        &\overset{\left(b\right) }{=} \frac{{\mit \Gamma} _{mn}A_{mn}K^{\theta}_{mn} abe^{-jk\left({r}_{0,mn}^{i}+{r}_{0,mn}^{r}\right) }}{2j\lambda {r}_{0,mn}^{r} {r}_{0,mn}^{i}} \times {\rm sinc}\left(\frac{kaA_{mn}^x}{2}\right)   {\rm sinc}\left(\frac{kbA^y_{mn}}{2}\right) 
         \label{dedede}
        \end{aligned} \tag*{(\arabic {equation})}
\end{equation}}
\rule[-7pt]{16.4cm}{0.05em}	
\end{figure*}

Figure~\ref{sinc} plots the factors ${\rm sinc}\left(\frac{kaA_{mn}^x}{2}\right)$ and ${\rm sinc}\left(\frac{kbA^y_{mn}}{2}\right)$ in Eq.~\ref{dedede} versus the angles $\theta^i_{mn}$, $\theta^r_{mn}$, $\phi^i_{mn}$, and $\phi^r_{mn}$, respectively. The aforementioned factors together with the leaning factor $K^{\theta}_{mn}$ are the specific radiation patterns of RIS unit elements referred in \cite{RIS_Path_Loss}.

{
\begin{remark}
    ${\rm sinc}\left(\frac{kaA_{mn}^x}{2}\right)$${\rm sinc}\left(\frac{kbA^y_{mn}}{2}\right)$$K^{\theta}_{mn}$ in Eq.~\ref{dedede} are the specific expressions of radiation patterns of RIS elements referred in \cite{RIS_Path_Loss}. The radiation patterns are mainly impacted by the physical sizes of elements as well as the angle of elevation and azimuth between RIS and Tx/Rx. 
\end{remark}}

{The electric fields from all individual elements in Eq.~\ref{dedede} will interfere at the Rx side and contribute to the final received electric field.} Thus, the sum of the reflected electric field from RIS, denoted by $E_r$, can be expressed as
\begin{align}
    {{E}}_r=\sum_{m = 1}^{M}\sum_{n = 1}^{N} {{E}}_{mn}^r.
\end{align}
The effective aperture of the receive antenna, denoted by $A_e$, can be written as $A_e=G_r\lambda^2/(4\pi)$. Thus, the received power reflected by RIS in NLoS case, denoted by $P_{r, {\rm NLoS}}$, can be expressed as Eq.~\ref{power1} according to the radiation law $P={\left\lvert E\right\rvert^2A_e}/\left({2\eta_0}\right)$. {As the RIS is typically deployed for NLoS scenarios to provide a virtual LoS link, we analyse the behavior of the proposed TPOSJ scheme under NLoS scenarios in \emph{Remark 3}.}

{\begin{remark}
The result in Eq.~\ref{power1} is in agreement with the existing work based on the concept of RCS when $\psi_{mn}^r=\pi$ \cite{Electromagnetic_Model} and also approaches the same maximum value as the derived results in \cite{RIS_Path_Loss} when the communication system is monostatic (i.e., the transmitter and receiver are at the same location which is perpendicular to the RIS). It can be seen from Eq.~\ref{power1} that the power gain is impacted by the leaning factor $K^{\theta}_{mn}$ and the angles including azimuth and elevation. When satisfying $\theta_{mn}^r=\theta_{mn}^i$ and $\phi_{mn}^r=\phi_{mn}^i+\pi/2$ (i.e., the specular reflection), it achieves the maximum value. In general scenarios that the RIS is distant from Tx and Rx, based on the \emph{Remark 1}, the performance of the TPOSJ scheme can be further improved by changing its facing orientation if its center is fixed. By rotating its facing direction, the optimal performance against position errors can be attained if $z_r^B=z_t^B$. Under such case, the errors in different directions are the same and the impacts of them on TPOSJ scheme are equivalent to AWGN channel. Additionally, when $z_r^B=z_t^B$, the leaning factor can also approach the maximum value with $\theta_{mn}^r=\theta_{mn}^i$ to improve the received power. 
\end{remark}}

\addtocounter{equation}{1}
\begin{figure*}[!bt]
    \begin{equation}
        \begin{aligned}
            P_{r,{\rm nLoS}}=\frac{P_t \varGamma^2a^2b^2}{16\pi^2} \left\lvert\sum_{m = 1}^{M}\sum_{n = 1}^{N} \frac{\sqrt{G_r\left(\theta_{mn}^{R}\right) G_t\left(\theta_{mn}^{T}\right)}  K^{\theta}_{mn} \varGamma_{mn} {{\rm sinc}\left(\frac{kaA_{mn}^x}{2}\right)   {\rm sinc}\left(\frac{kbA^y_{mn}}{2}\right)} }{{r}_{0,mn}^{r} {r}_{0,mn}^{i}} \right .\\
            \times \left .e^{-jk\left({r}_{0,mn}^{i}+{r}_{0,mn}^{r}\right) }\right\rvert^2 
            \label{power1}
        \end{aligned} \tag*{(\arabic {equation})}
    \end{equation}
\rule[-7pt]{16.4cm}{0.05em}	
\end{figure*}

When the direct link exists, the total received electric field can be modified as ${{E}}_{total}={{E}}_r+{{E}}_d$, where ${{E}}_d$ is the electric field from direct link. According to Friis transmission equation, ${{E}}_d$ can be expressed by
\begin{align}
    E_d=\sqrt{\frac{\eta_0 P_t G_t\left(\theta_d^T \right)}{2\pi r_d^2}}e^{-jkr_d}.
\end{align}
Consequently, the total received power in LoS case, denoted by $P_{r,{\rm LoS}}$, can be modified as Eq.~\ref{power2}.
{\begin{remark}
    It can be observed from Eqs.~\ref{power1} and \ref{power2} that the total reflected power is proportional to the square of area. Thus, to provide expected power gains in practice, the Tx and Rx will be more likely to be in the near-field. The maximal reflected power can be achieved when $\tau=0.5$. However, considering the position error, the value of $\tau$ should be properly designed to achieve the balance between performance and robustness to position error. Also, the relative orientation between RIS and Tx/Rx will have a huge impact on the performance of RIS even if the transmission distance is the same, which is related to the radiation pattern of element. 
\end{remark}}

\addtocounter{equation}{1}
\begin{figure*}[!bt]
    \begin{equation}
        \begin{aligned}
             P_{r,{\rm LoS}}=\frac{P_t}{16\pi^2}  \left\lvert  \sum_{m = 1}^{M}\sum_{n = 1}^{N} \frac{\sqrt{ G_t\left(\theta_{mn}^{T}\right)G_r\left(\theta_{mn}^{R}\right)}  ab  K^{\theta}_{mn}\varGamma_{mn}{{\rm sinc}\left(\frac{kaA_{mn}^x}{2}\right)   {\rm sinc}\left(\frac{kbA^y_{mn}}{2}\right)}}{{r}_{0,mn}^{r} {r}_{0,mn}^{i}} 
             \right.
                \\
            \phantom{=\;\;}\left. 
            \times e^{-jk\left({r}_{0,mn}^{i}+{r}_{0,mn}^{r}\right) }+\frac{\sqrt{G_t\left(\theta_d^T \right)G_r\left(\theta_d^R\right) }e^{-jkr_d}}{r_d}\right\rvert^2 
            \label{power2}
        \end{aligned} \tag*{(\arabic {equation})}
 \end{equation}
\rule[-7pt]{16.4cm}{0.05em}	
\vspace{-10pt}
\end{figure*}

\vspace{-5pt}
\section{Performance Evaluation} 
\begin{table}[!t] \scriptsize
    \centering
    \caption{{Simulation parameters}}
    \arrayrulecolor{black}
    \begin{tabular}{|c<{\centering}|c<{\centering}|c<{\centering}|}
    \hline
    \textbf{Parameter}& \textbf{Symbol}& \textbf{Value}\\
    \hline
    Frequency&  $f$ & 28 GHz\\
    {Transmit power}&  {$P$} & {30 dBm}\\
    {Number of antennas}&  {$B$} & {2}\\
    Tx antenna gain&  $G_T(0)$ & 15 dBi\\
    {Rx antenna gain}&  {$G_R(0)$} & {0 dBi}\\
    {Tx position}&  ${\rm T}^A$ & {$\left(10,10,15\right)$} \\
    {Rx position}&  ${\rm R}^A$ & {$\left(-10,10,1.5\right)$} \\
    {Noise power }&  $\sigma^2$&{ -90 dBm }\\
    Number of row &  $M$ & 80\\
    Number of column &  $N$ & 80\\
    Amplitude of reflection coefficient & $\varGamma$  & 0.9\\
    Element length&  $a$ & $\frac{\lambda}{2}$\\
    Element width&  $b$ & $\frac{\lambda}{2}$\\
    \hline
\end{tabular}
\vspace{-10pt}
\end{table}
In this section, we evaluate the performance of our proposed scheme. The simulation parameters are summarized in Table II unless specified otherwise. 

\begin{figure*}[t]
    \centerline{\includegraphics[width=14cm]{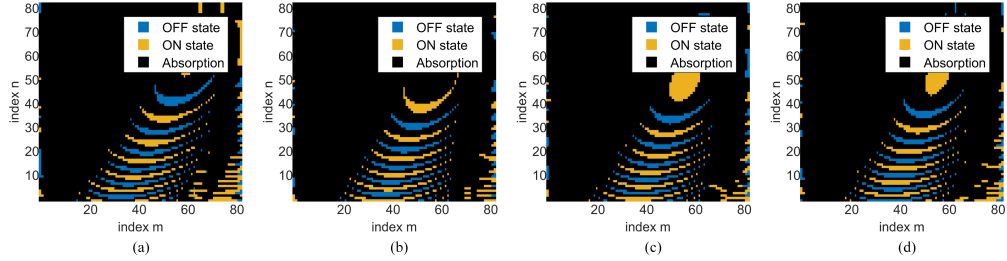}}
    \vspace{-5pt}
    \caption{The RIS ON/OFF states configuration versus the columns and rows. (a) The position information of Rx is accurate. (b) The position information of Rx is accurate ($e_x=0.1$ m). (c) The position information of Rx is accurate ($e_y=0.1$ m). (d) The position information of Rx is accurate ($e_z=0.1$ m). }
    \label{onoff}
    \vspace{-10pt}
\end{figure*}

Figure \ref{onoff} depicts one implementation example of the RIS configuration matrix under the case of accuracy and inaccuracy, respectively. As depicted in Fig.~\ref{onoff}(a), in the accurate case, the RIS elements in the same state form one part of the ellipse and compose a zonary structure, which conforms to the properties of the Fresnel zone. {The impacts of individual error components on the configuration matrix are depicted in Figs. \ref{onoff}(b), \ref{onoff}(c) and \ref{onoff}(d), where the variance of all error components are 0.1 m. In this example, the performance under $e_x$ is more severe than that of $e_y$ and $e_z$ as it lead to the opposite phase manipulation, degrading the received power. Additionally, the phase configuration is not sensitive to $e_z$, which is consistent with our analysis in Section III-D. These position errors lead to the wrong judgement of the element state, thus degrading the performance of the TPOSJ scheme. }

\begin{figure}[t]
    \centerline{\includegraphics[width=9cm]{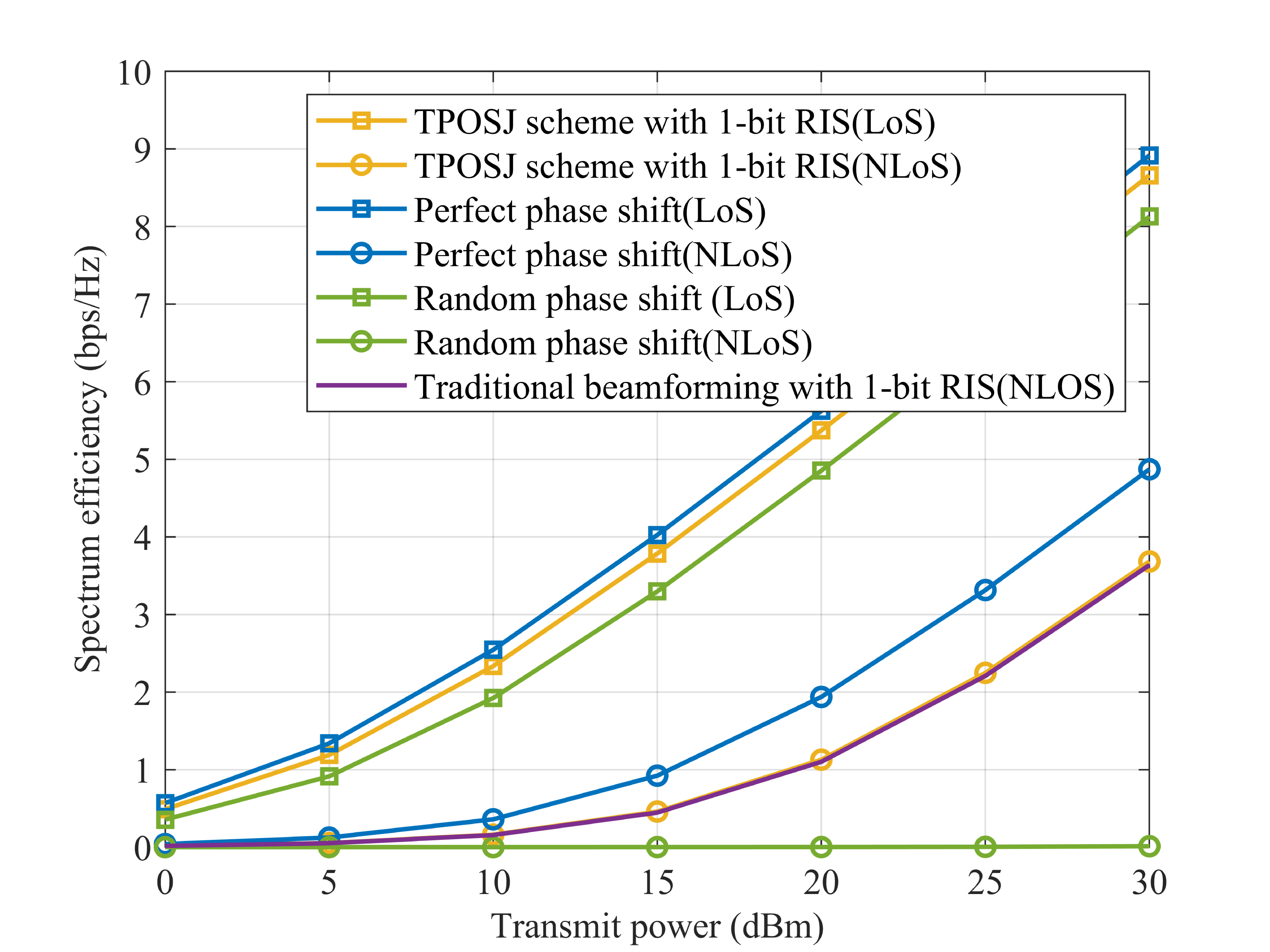}}
    \caption{{The spectrum efficiency versus the transmit power in dBm.}}
    \label{sp_compare}
    \vspace{-10pt}
\end{figure}

{Figure \ref{sp_compare} plots the spectrum efficiency of our proposed TPOSJ scheme ($\tau=0.5$) versus transmit power from 0 dBm to 30 dBm under LoS and NLoS cases, compared with the traditional near-field RIS-assisted beamforming\cite{RIS_Path_Loss}. Each case is divided into three conditions distinguished by the condition of phase shift including the random phase shift, the 1-bit phase shift, and the perfect phase shift (continuous phase shift). In the simulation, the radiation pattern of the transmit and receive antenna is set as ${\rm cos}^q$ function.} Thus, the gain of the antenna towards a certain direction can be computed as follows:
\begin{equation}
    \begin{aligned}
        G\left(\theta\right) =\frac{4\pi {\rm cos}^q\left(\theta\right) }{\int_0^{2\pi}\int_0^{\frac{\pi}{2}}{\rm cos}^q\left(\theta\right)d\theta d\phi} ,
    \end{aligned}
\end{equation}
{where $q$ depends on the boresight gain satisfying $G(0)=2(q+1)$\cite{Localization3}, $\phi$ and $\theta$ are the corresponding azimuth and elevation angles, respectively. The value of $q$ is 14.8 when the boresight gain of transmit antenna is set as 15 dBi in 28 GHz\cite{antenna_dbi}, where the main lobe is toward the center of RIS array. As shown in Fig.~\ref{sp_compare}, the spectrum efficiency of our proposed TPOSJ scheme with $\tau=0.5$ is equivalent to that of the benchmark case, which locates between the perfect phase shift case and the random phase shift case in both LoS and NLoS cases. In the NLoS case, our proposed scheme shows a significant gain compared with random phase shift, thus revealing the importance of phase manipulation. When the LoS path exists, the gain of RIS is small since the RIS with the size of $0.8 {\rm m}\times 0.8 {\rm m}$ can only provide limited reflected power.}

\begin{figure}[t]
    \centerline{\includegraphics[width=9cm]{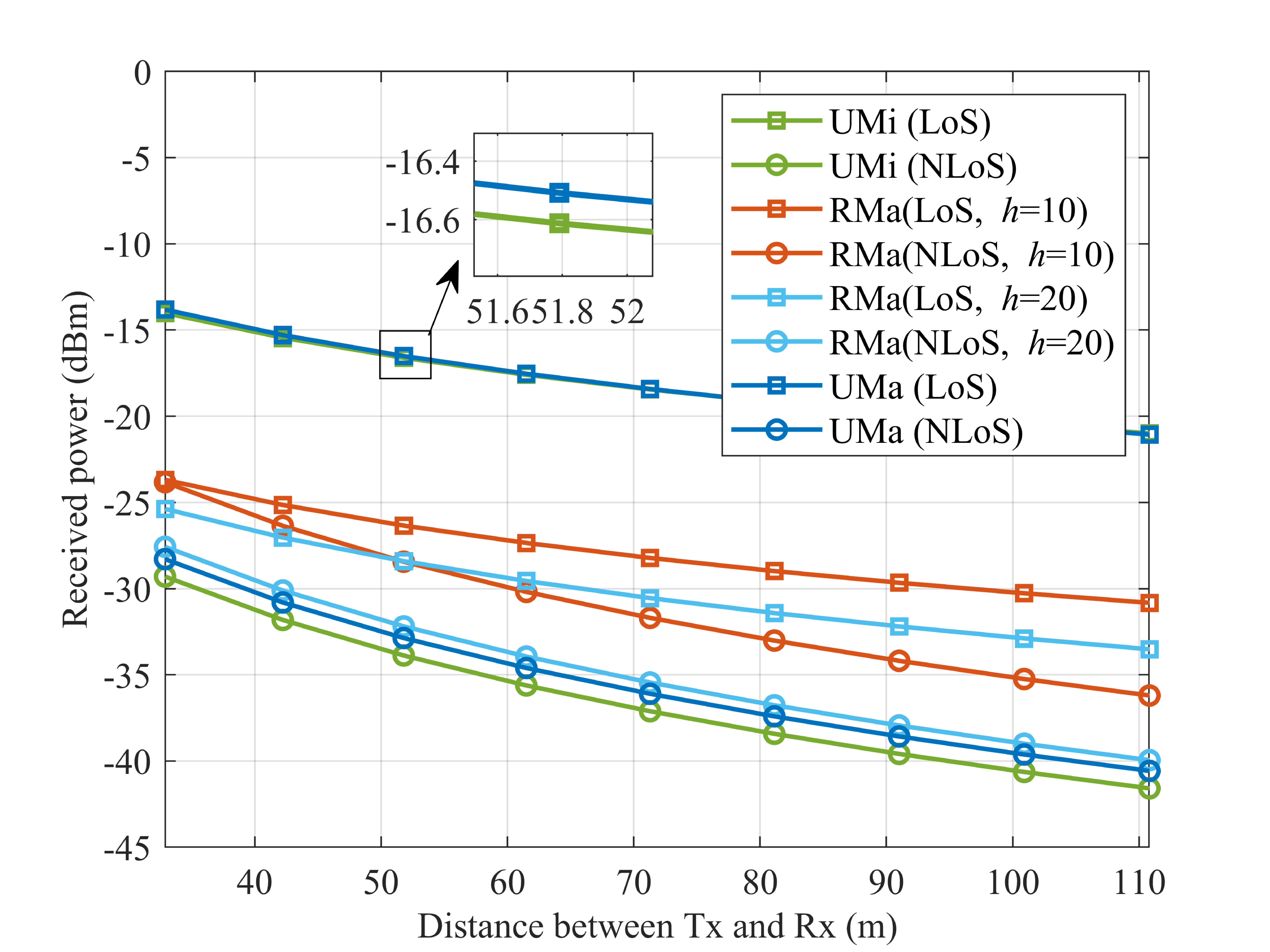}}
    \caption{{The received power versus the transmission distance under different conditions.}}
    \label{sp_shape_surf}
    \vspace{-10pt}
\end{figure}
{Figure \ref{sp_shape_surf} shows the received power of the proposed scheme against the transmission distance under varying different conditions including urban microcell (UMi), Urban Micro (UMa), and rural macrocell (RMa) with different average building heights ($h$) in both LoS and NLoS scenarios. All parameters corresponding to the above path loss model are determined by 3GPP TR 38.901 version 16.1.0 Release 16 (Table 7.4.1-1)\cite{ETSI}. The average building height $h$ is set as 10 m and 20 m (high rise building scenario) for RMa scenario to evaluate the impact of high rise buildings. As shown in Fig.~\ref{sp_shape_surf}, as the RIS can only provide limited reflected gain, there is a significant gap of the received power between the NLoS and LoS conditions. Therein, the gaps in UMi and UMa scenarios are larger than that of the RMa case. Meanwhile, the received power for UMi and UMa scenarios behave similarly with only slight differences. For the RMa scenario, the received power decreases as the average building height $h$ increases, since the probability that the dominant propagation paths are blocked increases as $h$ increases.
}

\begin{figure}[t]
    \centerline{\includegraphics[width=9cm]{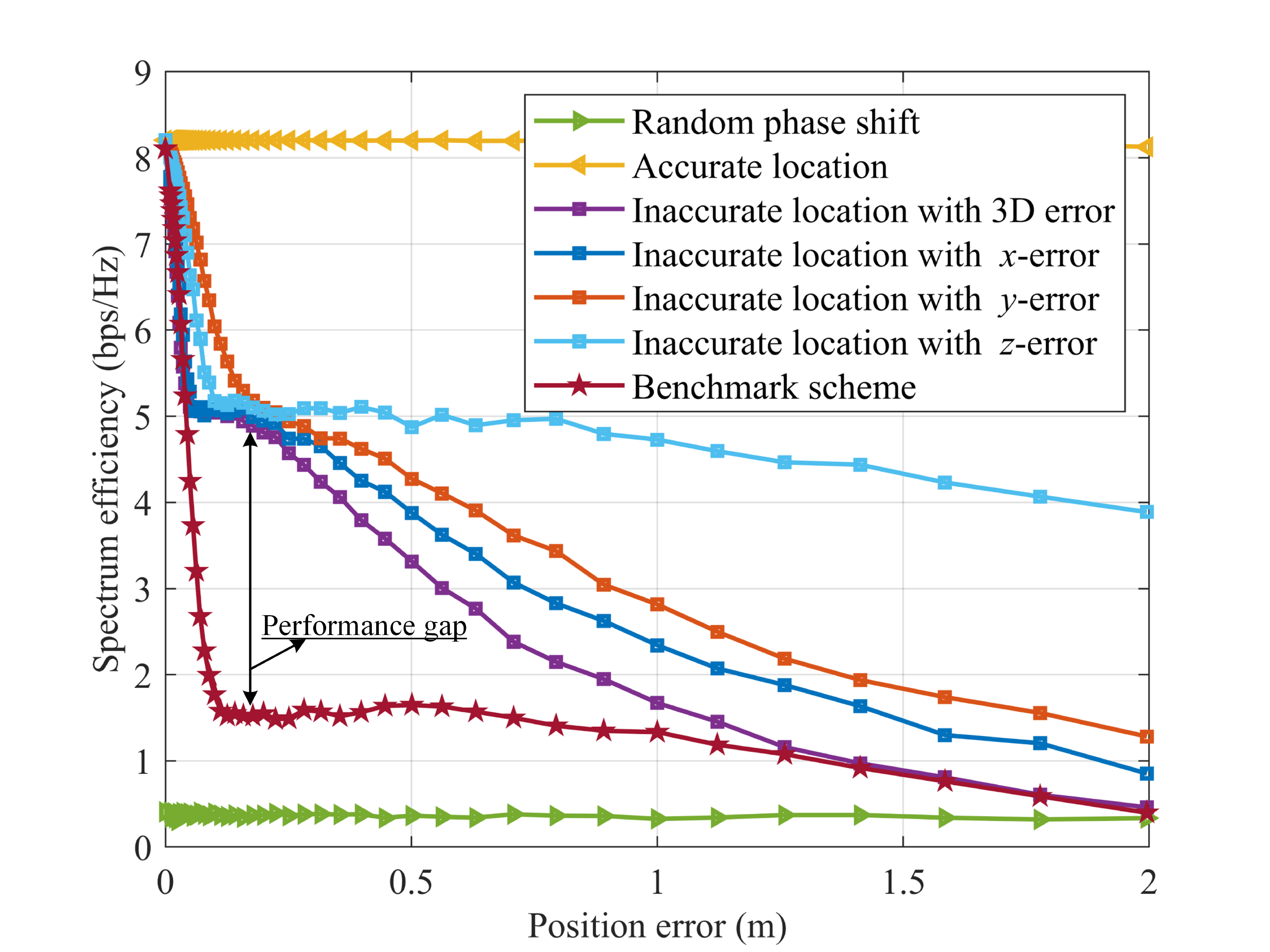}}
    \vspace{-10pt}
    \caption{{The spectrum efficiency versus the position error.}}
    \label{error_on_sp}
    \vspace{-10pt}
\end{figure}

{Figure \ref{error_on_sp} plots the impacts of 3D position error and its individual components corresponding to different axes on spectrum efficiency in NLoS case, considering the traditional beamforming as the benchmark scheme. As shown in Fig.~\ref{error_on_sp}, with the increase of position error, the spectrum efficiency of our proposed scheme first rapidly degrades in low error regime, and later slowly degrades until approaching the performance of random phase shift. The spectrum efficiency of the benchmark scheme decreases more seriously as the position error increases, which also finally approaches the performance of random phase shift. Meanwhile, our proposed TPOSJ scheme is robust to position errors to a certain extent while the performance of traditional beamforming degrades rapidly as the position error increases. This is because our proposed TPOSJ scheme leverages the property of Fresnel zone, which can capture the property of the wavefront on RIS. In addition, the declines of spectrum efficiency versus $e_x$ and $e_y$ are more severe than that of on $e_z$, which reveals that the position error on $z$-coordinate has a slight impact on spectrum efficiency compared with other components. Furthermore, the impact of error on $x$-coordinate is more severe than that of on $y$-coordinate, which is consistent with our theoretical analysis. }

\begin{figure}[t]
    \centerline{\includegraphics[width=9cm]{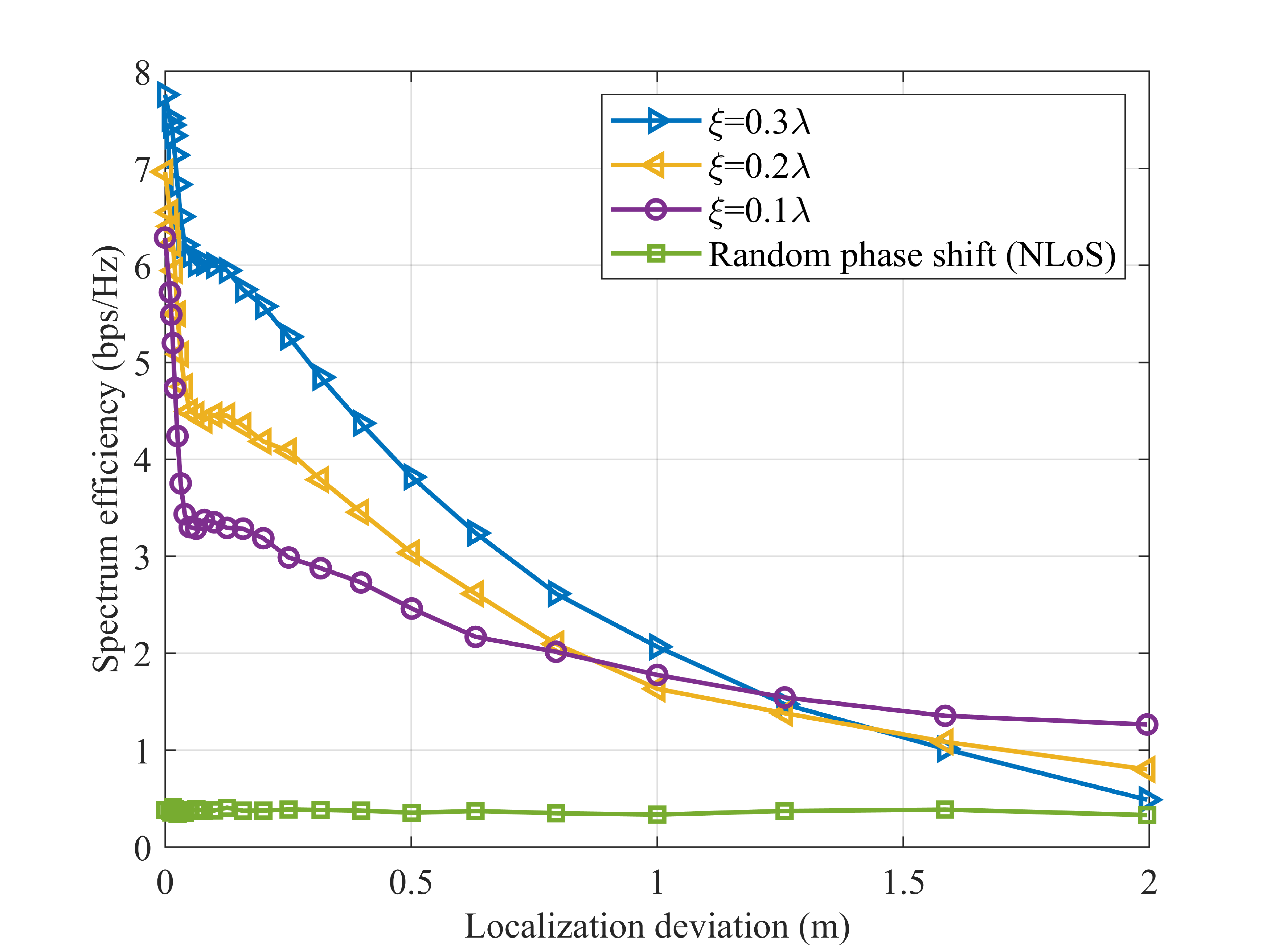}}
    \caption{{The spectrum efficiency versus position error in NLoS environment when $\xi=0.1\lambda, 0.2\lambda, 0.3\lambda$.}}
    \label{xi_tu}
    \vspace{-10pt}
\end{figure}
{Figure \ref{xi_tu} shows the spectrum efficiency versus position error in NLoS environment when $\xi=0.1\lambda, 0.2\lambda, 0.3\lambda$, respectively. As shown in Fig.~\ref{xi_tu}, the spectrum efficiency in the absence of position error increases as $\xi$ increases, which is consistent with Eq.~(17) since there are fewer elements configured as absorption mode as the threshold increases. This will increase the maximum achievable rate. In low position error region, the performance with higher value of $\xi$ surpasses that of with lower value. That is, increasing the judgement threshold can improve the performance in low position error region. Nevertheless, as $\xi$ decreases, the rate of decline slows down and the gentle interval extends. This is because the reflected signals from these elements in reflection mode behave strong constructive superposition, which is less impacted by position error. Conversely, in high position error regime, the performance with lower value of $\xi$ surpasses that of with higher value.  Decreasing the judgement threshold can improve the performance in high position error region. Thus, the absorption mode can be used to improve the robustness for the scenario where the localization accuracy is coarse or Tx is on moving with mobility. The tradeoff between maximum achievable rate and robustness to position error can be chosen in view of the extent of practical position error. }

\section{Conclusions}
In this paper, we tackled the problem of how to configure the phase shift without CSI in the near-field RIS-assisted communications. In particular, we designed the impedance based discrete RIS elements with joint absorption mode and reflection mode. {Then, we developed CSI-free position-aided phase configuration scheme with low complexity, which combines the property of Fresnel zone with position information to alleviate the dependence on channel estimation in the near-field RIS-assisted communications.} Also, the corresponding frame structure was designed and the impacts of mobility and position error were analyzed to verify the feasibility of our proposed scheme. Finally, based on the proposed scheme, we derived the power flow of RIS by the Helmholtz-Kirchhoff integral theorem. The impacts on spectrum efficiency versus different sets were analyzed in simulation including the transmit power, the shape of RIS array, the position error, and the judgement threshold. Numerical results showed that our proposed scheme can manipulate the ON/OFF state intelligently without complex CSI. {Also, compared with traditional beamforming scheme, our proposed scheme is less impacted by position errors in specific directions, thus verifying the potential application of our proposed scheme. For the extension of multiple-antenna Rx, we intend to defer a comprehensive exploration of this matter to future research endeavors.}
\vspace{-5pt}

\begin{appendices}
\section{Proof of Proposition 1}
According to Helmholtz-Kirchhoff integral theorem, the reflected electric field at the observation point can be derived by the electric field and its derivative on any closed surface surrounding the observation point. As shown in Fig.~\ref{Huygens}, the electric field at Rx can be expressed as follows:
\begin{equation}
   {E}_{mn}^r=\frac{1}{4\pi}\oiint_{\varSigma_1+\varSigma_2+S_{mn}}\left[{\frac{\partial {E}}{\partial \mathbf{ {n}}}}G\left(\mathbf{ {r}}_{mn}^{i}\right)  - {E}\frac{\partial G\left(\mathbf{ {r}}_{mn}^{r}\right)}{\partial \mathbf{ {n}}}\right] d\sigma,
\end{equation}
where ${E}$ is the electric field at the boundaries.

Since the boundary $\varSigma_1$ is a hemisphere with an infinite radius, the electric field $E$ at this boundary can be ignored. In addition, there is no contribution to the electric field of observation point from the boundary $\varSigma_2$. Thus, we omit the electric field at the boundary $\varSigma_2$ as we only concentrate on the reflected field from the RIS element. Thus, the reflected electric field can be simplified as follows:
\begin{equation}
    {E}_{mn}^r=\frac{1}{4\pi} \iint_{S_{mn}}\left[{\frac{\partial {E}_S}{\partial \mathbf{ {n}}}}G\left(\mathbf{ {r}}_{mn}^{i}\right)  - {E}_S\frac{\partial G\left(\mathbf{ {r}}_{mn}^{r}\right)}{\partial \mathbf{ {n}}}\right] d\sigma,\label{reflected}
\end{equation}
where ${E}_S$ is the electric field at the element surface. As the polarization of reflected wave is properly matched, the electric field ${E}_S$ can be expressed as follows:
\begin{align}
     {E}_S={\mit \Gamma} _{mn}  {E}_{inc}(z=0^+) , \label{es}
\end{align}
where $ {E}_{inc}(z=0^+)$ is the incident electric field at the surface of $S_{mn}$. According to Green's theorem, $ {E}_{inc}(z=0^+)$ can be expressed by 
\begin{align}
     {E}_{inc}(z=0^+)= A_{mn}G( \mathbf{r}_{mn}^{i})=\frac{A_{mn}  {\rm exp}\left(-jk {r}_{mn}^{i} \right)}{ {r}_{mn}^{i}}.\label{reflect}
\end{align}
In addition, the directional derivative of Green's function can be expressed as follows:
\begin{equation}
\begin{aligned}
    \frac{\partial G\left( \mathbf{r}\right)}{ \mathbf{n}}=&-\cos( \mathbf{r}, \mathbf{n})\left(jk-\frac{1}{\mathbf{r}}\right)\frac{e^{-jk{r}}}{{r}} \\ 
    \overset{\left(a\right) }{\approx}& -{jk}\cos( \mathbf{r}, \mathbf{n})\frac{e^{-jk{r}}  }{{r}}, \label{green}
\end{aligned} 
\end{equation}
where the approximation in $\left(a\right)$ holds as $1/r$ $\ll k$ holds{\footnote{From a theoretical perspective, as the far-field condition of individual element always holds under general communication distance, the wave's curvature caused by $1/\mathbf{r}$ is negligible so that the approximation holds. On the other hand, the wavelength $\lambda$ is at most 10 mm in high frequency bands (i.e, from 30 GHz to 3 THz). The typical transmission distance is usually tens or hundreds of meters and its reciprocal is far less than one. Under such condition, the inequality $1/r$ $\ll k$ always holds so that the approximation is reasonable.}}. Substituting Eqs.~\eqref{es}, \eqref{reflect}, and \eqref{green} into Eq.~\eqref{reflected}, we can derive the reflected electric ${E}_{mn}^r$ as follows:
\begin{align}
    &{E}_{mn}^r=\frac{jk{\mit \Gamma} _{mn}A_{mn}}{4\pi}\times  \\
    &\iint_{S_{mn}}\left\{G\left(\mathbf{r}_{mn}^{i}\right) G\left(\mathbf{r}_{mn}^{r}\right) \left[\cos\left( \mathbf{r}_{mn}^{r}, \mathbf{n}\right) -\cos\left( \mathbf{r}_{mn}^{i}, \mathbf{n}\right)\right] \right\} d\sigma.
\end{align}
\vspace{-5pt}

\section{Proof of Theorem 2}
The geometry of Tx and Rx of the $A$ coordinate systems is shown in Fig.~\ref{fresnel_geo}. The extension lines of $ \rm{TR} $ and $ \rm{T_\bot R_\bot}$ cross at point $\rm O$. As shown in Figs.~\ref{fresnel_geo}, we can derive the corresponding relation including $\left\lvert \rm{RP}\right\rvert=d_1$, $\left\lvert   \rm{PT}\right\rvert=d_2$, $\left\lvert   \rm{RT}\right\rvert=d$, $\left\lvert   \rm{RR_\bot}\right\rvert=z_r^B$, $\left\lvert   \rm{TT_\bot}\right\rvert=z_t^B$, and $\left\lvert   \rm{R_\bot T_\bot}\right\rvert=x_t^B-x_r^B$. The lengths of the red lines, denoted by $ \rm{PQ}$ and $ \rm{R_\bot Q}$, are the targets. Thus, we can derive
\begin{align}
    \begin{cases}
     &\frac{\left\lvert   \rm{RR_\bot}\right\rvert }{\left\lvert   \rm{TT_\bot}\right\rvert}=\frac{\left\lvert   \rm{OR}\right\rvert}{\left\lvert   \rm{OR}\right\rvert+\left\lvert   \rm{RT}\right\rvert};\\
     &\frac{\left\lvert   \rm{RR_\bot}\right\rvert }{\left\lvert   \rm{TT_\bot}\right\rvert}=\frac{\left\lvert   \rm{OR}_\bot\right\rvert}{\left\lvert   \rm{OR}_\bot\right\rvert+\left\lvert   \rm{RT}_\bot \right\rvert};\\
     &\frac{\left\lvert   \rm{PQ}\right\rvert}{\left\lvert   \rm{OR}\right\rvert +\left\lvert   \rm{RP}\right\rvert}=\frac{\left\lvert   \rm{RR_\bot}\right\rvert}{\left\lvert   \rm{OR}_\bot\right\rvert};\\
     &\frac{\left\lvert   \rm{OR}_\bot\right\rvert+\left\lvert   \rm{R_\bot Q}\right\rvert}{\left\lvert   \rm{OR}\right\rvert}=\frac{\left\lvert   \rm{PQ}\right\rvert}{\left\lvert   \rm{RR_\bot}\right\rvert}.
    \end{cases} \label{geo}
\end{align}
Thus, the length of $  \rm{PQ}$ and $  \rm{R_\bot Q}$ can be derived from Eq.~\eqref{geo} as 
\begin{align}
    \left\lvert   \rm{PQ}\right\rvert=\frac{ \left\lvert \rm{RR_\bot}\right\rvert  \left\lvert \rm{PT}\right\rvert + \left\lvert \rm{TT_\bot}\right\rvert  \left\lvert \rm{RP}\right\rvert }{ \left\lvert \rm{TT_\bot}\right\rvert - \left\lvert \rm{RR_\bot}\right\rvert } \label{length1}
\end{align}
and 
\begin{equation}
    \begin{aligned}
        \left\lvert   \rm{R_\bot Q}\right\rvert=\frac{ \left\lvert   \rm{PQ}\right\rvert\left\lvert \rm{RT}\right\rvert }{  \left\lvert \rm{R_\bot T_\bot}\right\rvert } -\frac{\left\lvert \rm{RR_\bot}\right\rvert }{\left\lvert \rm{TT_\bot}\right\rvert - \left\lvert \rm{RR_\bot}\right\rvert},\label{length2}
\end{aligned}
\end{equation}
respectively.
{Substituting the corresponding lengths into Eqs.~\eqref{length1} and \eqref{length2}, the lengths of $\rm{PQ}$ and $\rm{R_\bot Q}$ can be expressed as
\begin{align}
    \left\lvert   \rm{PQ}\right\rvert=\frac{z_r^Bd_2+z_t^Bd_1}{\left\lvert z_t^B-z_r^B\right\rvert }
\end{align}
and 
\begin{align}
    \left\lvert  \rm{R_\bot Q}\right\rvert=\frac{\left(z_r^Bd_2+z_t^Bd_1\right)d-z_r^B\left(x_t^B-x_r^B\right)^2}{\left\lvert z_t^B-z_r^B\right\rvert \left\lvert x_t^B-x_r^B\right\rvert },
\end{align}
respectively.}

\end{appendices}

\bibliography{References}
\bibliographystyle{ieeetr}

\end{document}